\documentclass[graybox, nosecnum]{svmult}

\usepackage{mathptmx}       % selects Times Roman as basic font
\usepackage{helvet}         % selects Helvetica as sans-serif font
\usepackage{courier}        % selects Courier as typewriter font
\usepackage{type1cm}        % activate if the above 3 fonts are
\usepackage{makeidx}         % allows index generation
\usepackage{graphicx}        % standard LaTeX graphics tool
\usepackage{multicol}        % used for the two-column index
\usepackage[bottom]{footmisc}% places footnotes at page bottom
\usepackage{hyperref}        %for hyperlinks
\usepackage{soul}            % for high-lighting of text

\usepackage{bm}% bold math
\usepackage{xcolor}
\usepackage{amsmath}    % need for subequations
\makeindex             % used for the subject index
\begin{document}
%\tableofcontents{}
\title*{Model for independent particle motion}
\author{A. V.  Afanasjev \thanks{corresponding author}}
\institute{
A. V. Afanasjev \at Department of Physics and Astronomy, Mississippi
State University, MS 39762, USA, \email{Anatoli.Afanasjev@gmail.com}}
%\and Second Author \at Institute 2, Address of Institute 2 \email{name2@email.address}}
\maketitle
\abstract{Independent particle model in nuclear physics assumes 
that the nucleon in the nucleus moves in the average (mean field) potential 
generated by all other nucleons. This chapter gives a short overview of basic
features of the independent particle motion in atomic nuclei and its theoretical
realization in the framework of shell models for spherical, deformed and 
rotating nuclei as well as in more sophisticated approaches such as  
microscopic+macroscopic model and density functional theories. Independent
particle motion of nucleons leads to global and single-particle consequences.
The global ones manifest themselves in the shell structure and its consequences 
for global structure of nuclear landscape, the existence of superheavy nuclei and 
the superdeformation at high spin are briefly reviewed. The latter shows itself in 
the single-particle properties such as energies, alignments and densities; their 
manifestations are illustrated on specific examples.
}

%%%%%%%%%%%%%%%%%%%%%%%%%
\section{Introduction}
%%%%%%%%%%%%%%%%%%%%%%%%%

    The basic idea of the independent particle model is that nucleon (proton or neutron) 
moves  in an  average (mean field) nuclear potential and this motion is independent of motion of 
other nucleons.  This allows to replace the complicated solution of the problem of $A$ interacting 
particles ($A$ is total number of nucleons in the nucleus) with much simpler problem of $A$ 
noninteracting particles in the mean field potential.  This mean field potential can be either designed 
in a phenomenological way or calculated fully self-consistently from effective interaction. As discussed 
in this chapter, many properties of the nuclei have been described with high accuracy within such 
frameworks and many new phenomena (for example, superheavy nuclei and superdeformation at 
high spin) have been predicted theoretically and later observed experimentally. Independent particle 
model in its different realizations provides also the basis [mean field]  for the models which take into 
account residual interactions such as pairing, vibrations,  particle-vibration coupling  etc.  These interactions 
are typically included either by adding respective terms to the version of  independent particle model or by the 
modification of the formalism. 

The basic features of independent particle model, its realization in the shell model variants, 
microscopic+macroscopic model and covariant density functional theory will be discussed in 
this chapter. In addition, some manifestations of the independent particle motion emerging from 
the shell structure and single-particle properties will be considered. In no way, this chapter should 
be considered as "all-inclusive": it only scratches the surface of huge body of experimental and 
theoretical results.  More comprehensive  and detailed reviews on specific phenomena/theoretical 
approaches are quoted in  respective sections. In addition, some aspects of the independent particle 
motion are discussed in the books \cite{Bohr1975,RS.80,Casten-book,NilRag-book}.

%%%%%%%%%%%%%%%%%%%%%%%%%
\section{Independent particle model}
%%%%%%%%%%%%%%%%%%%%%%%%%

   The solution of the many-body nuclear problem for a nucleus consisting of $A$ 
nucleons ($Z$ protons and $N$ neutrons)  requires the solution of the eigenvalue problem
\cite{Bohr1975,RS.80}
\begin{eqnarray}
H \Psi_{\alpha} (\vec{r},\sigma, \tau) = E_{\alpha} \Psi_{\alpha} (\vec{r},\sigma, \tau).
\label{Hamilt} 
\end{eqnarray}

  The Hamiltonian of the system could be either in non-relativistic (Schr\"{o}dinger equation)  or
relativistic (Dirac equation) forms.  It contains the kinetic and potential energy contributions from
each nucleon
\begin{eqnarray}
H=\sum_{i=1}^{A} \frac{\hbar^2}{2m_i} \nabla_i^2 + \sum_{i\neq j} V_{ij}\,\,, 
\end{eqnarray} 
and includes two-body interaction between $i$- and $j$-th nucleons. 

   The total wave function $\Psi_{\alpha} (\vec{r},\sigma, \tau)$ of  the nucleus 
in the state with quantum numbers $\alpha$ has to be expressed in terms of the 
ones of the individual nucleons  $\psi_i(\vec{r}_i,\sigma_i, \tau_i)$. Here $\vec{r}$, $\sigma$ 
and $\tau$ are position, spin and isospin variables, respectively. For simplicity, in further 
discussion $\vec{r}_i$ will represent all independent variables of the $i$-th nucleon.
Total wave function of the system of the $A$ particles has to be antisymmetric with 
respect of the exchange of the coordinates of two particles. Thus, a many-body
state is written in the form of the Slater determinant \cite{RS.80}
\begin{eqnarray} 
\Psi_{\alpha} (\vec{r}_1, \vec{r}_2, . . ., \vec{r}_A)
 = \frac{1}{\sqrt{A!}} \rm{det}
\left|  \begin{array}{cccc} 
\psi_1(\vec{r}_1) &   \psi_1(\vec{r}_2) & ... &   \psi_1(\vec{r}_A) \\
\psi_2(\vec{r}_1) &   \psi_2(\vec{r}_2) & ... &   \psi_2(\vec{r}_A) \\
 ...  &   ...  & ... &  ... \\
 \psi_A(\vec{r}_1) &   \psi_A(\vec{r}_2) & ... &   \psi_A(\vec{r}_A) 
\end{array} 
\right|. 
\end{eqnarray}
The factor $\frac{1}{\sqrt{A!}}$ is due to normalization. The choice of the single-particle 
wavefunctions defines the type of the many-body state. It is important to remember that 
the single-particle spectrum is infinite one. Thus, in practical applications respective 
Hilbert space is truncated most frequently based on energy considerations. 

   In nuclear systems it is possible to recast Eq.\ (\ref{Hamilt}) in the form
\begin{eqnarray}   
H = \sum_{i=1}^{A} h(\vec{r}_i) + \sum_{i\neq j=1}^{A} \tilde{V}(\vec{r}_i, \vec{r}_j),
\label{Hamilt-mod}
\end{eqnarray}    
where  $\tilde{V}(\vec{r}_i, \vec{r}_j)$ is the residual two-body interaction and $h(\vec{r}_i)$ 
is the single-particle hamiltonian. This recast essentially means that the part of original 
two-body interaction $V_{ij}$ is moved to the single-particle hamiltonian 
$h(\vec{r}_i)$. The latter can be chosen in such a way that the contribution of
residual interaction becomes rather small or even negligible. 

{\it Independent particle model} corresponds to the situation when second term in 
Eq.\ (\ref{Hamilt-mod}) is neglected. Thus, the nuclear Hamiltonian is a sum of 
single-particle terms
\begin{eqnarray} 
H= \sum_{i=1}^{A} h(\vec{r}_i),
\label{sum-sp-ham} 
\end{eqnarray}
and eigenvalue problem is reduced to 
\begin{eqnarray}
h(\vec{r}_i) \psi_k(\vec{r}_i) = e_k \psi_k(\vec{r}_i),
\end{eqnarray} 
where $e_k$
%$\varepsilon_k$ 
stand for the single-particle energy.  

 Independent particle model allows to reduce the nuclear many-body problem
from complicated two-body treatment to much simpler one-body one. It also relies 
on the use of the single-particle potentials. 
Independent particle model corresponds to the 
description of a nucleus in terms of noninteracting particles in the orbitals of these 
single-particle potentials which itself are produced by all the nucleons. Thus, 
nucleons move essentially free in these potentials. In this model, 
the ground state is formed by filling all the single-particle states located 
below the Fermi level. The excitation of the particle from occupied state
below this level to an empty one above it leads to an excited state. 
This process is called as a particle-hole excitation.

  The justifications for an independent particle model have been discussed in a number 
of publications \cite{Bohr1975,Casten-book}. The mean free path between 
collisions of the constituent nucleons is large compared to the average distance between 
them and  in some cases it could be larger than the dimensions of the nucleus.  As a 
consequence, the interactions between nucleons contribute mostly to the smoothly  
varying average potential in which the particles moves independently. One can arrive  to 
the same conclusion by taking into account the fact that the nucleons occupy only 
approximately 1\% of the volume of the nucleus. This estimate follows from the
consideration of nucleons as hard spheres of radius $c\approx 0.5$ fm. This value
of $c$ corresponds to the half of the closest distance which two nucleons can approach 
each other due to infinite repulsion (see discussion in Sec. 2-5b of Ref.\ \cite{Bohr1975}). 
For such value of $c$, nuclear matter behaves close to the free gas of nucleons with 
very rare head-on collisions of nucleons. The Pauli principle and the weakness
of strong nuclear interaction when compared with characteristic kinetic energies of
nucleons inside the nucleus are other contributing factors for the justification of
independent particle model \cite{Bohr1975,Casten-book}.

%%%%%%%%%%%%%%%%%%%%%%%%%
\section{Spherical shell model}
%%%%%%%%%%%%%%%%%%%%%%%%%

   The discussion of previous section clearly illustrates the need for the introduction
of the single-particle potential. This one-body potential 
can be introduced either in phenomenological or in a fully  
self-consistent ways. In the former case, one deals  with phenomenological Nilsson, 
Woods-Saxon or folded Yukawa potentials. Self-consistent
single-particle potentials are formed as a result of the solution of the many-body
nuclear problem within non-relativistic and relativistic density functional theories (DFTs);
note that these potentials are not treated separately from the rest of the nuclear many-body
problem.

%%%%%%%%%%%%%%%%%%%%%%%%%%%%%%%%%%%%%%%%% 
\begin{figure}[t!]
\centering
\includegraphics[width=\linewidth]{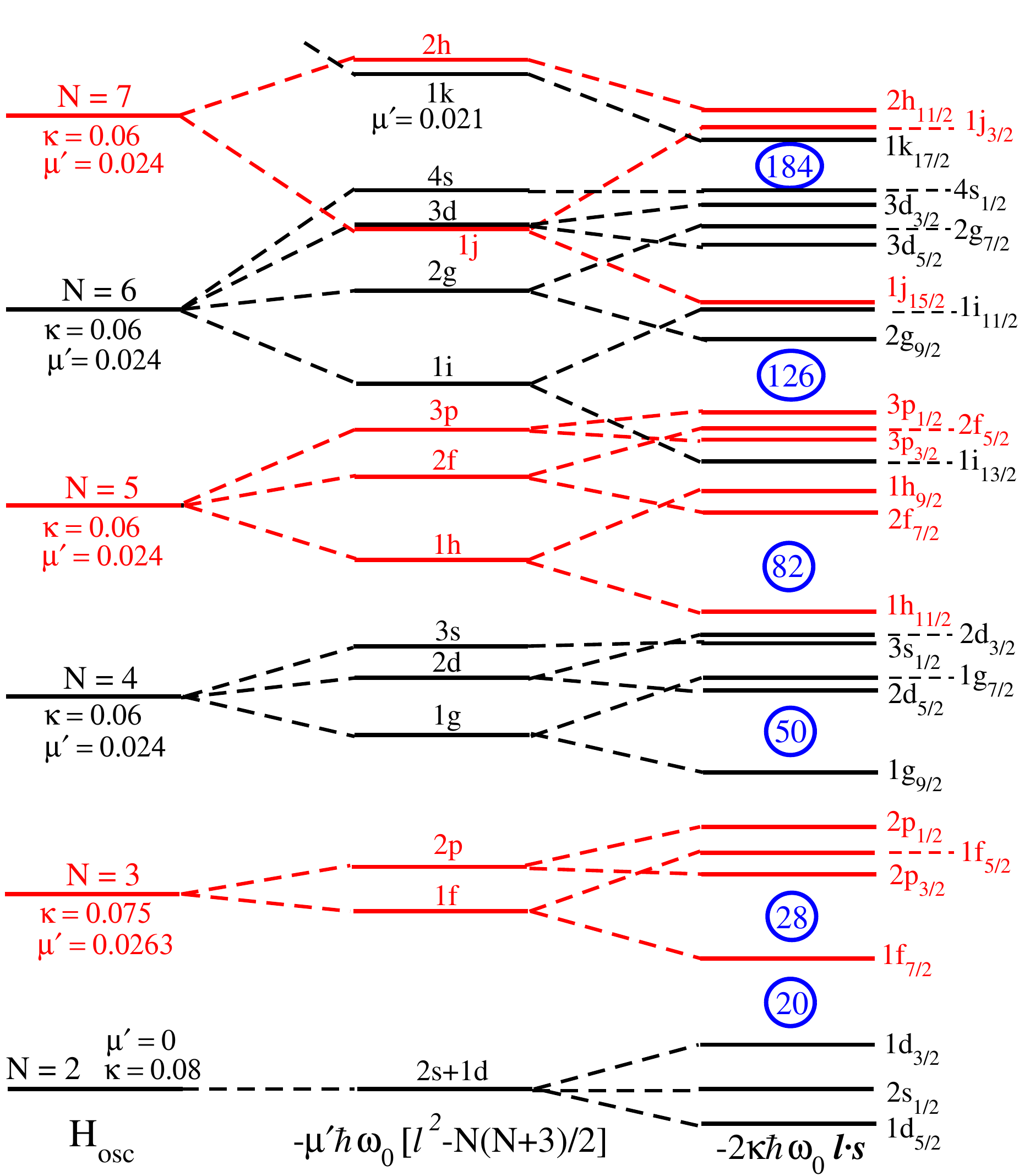}
\caption{The sequential build-up of realistic nucleonic potential. The left column 
shows the single-particle states of the pure harmonic oscillator. The employed parameters
$\kappa$ and $\mu'$ of the MO potential are displayed: note that they 
are different for the different $N$-shells. The modifications introduced by Eq.\ (\ref{V-corr-1}) 
are shown in the middle column. Finally, the right column shows the 
impact of spin-orbit interaction on the energies of single-particle states and their ordering.
Particle numbers corresponding to spherical shell closures are encircled. Black and
red colors are used for positive and negative parity states, respectively.  The figure is based 
on the  results presented in Fig. 6.3 of  Ref.\ \protect\cite{NilRag-book}.
\label{s-p-spher}
}
\end{figure}
%%%%%%%%%%%%%%%%%%%%%%%%%%%%%%%%%%%%%%%%% 

   To illustrate  the major physics aspects behind these potentials and
for pedagogical reasons,
% I will consider  
% harmonic oscillator (HO) potential and its modifications relevant for nuclear physics
% problems. 
the harmonic oscillator (HO) potential and its modifications relevant for nuclear physics
problems will be considered here. 
The single-particle hamiltonian of the nucleon with mass $m$ entering 
into Eq.\ (\ref{sum-sp-ham}) is given by
\begin{eqnarray} 
h(\vec{r}) = - \frac{\hbar^2}{2m} \nabla_i^2 + V(\vec{r}),
\end{eqnarray} 
where the central potential
\begin{eqnarray} 
V(\vec{r})= \frac{1}{2} m\omega_{\,0} r^2
\label{central}
\end{eqnarray} 
is pure HO potential representing one-body (mean) potential (field).
This hamiltonian in spherical coordinates is given by 
\begin{eqnarray} 
h(\vec{r}) = - \frac{\hbar^2}{2m} \frac{1}{r} \frac{\partial^2}{\partial r^2} r  + \frac{l^2(\theta,\phi)}{2mr^2} +V(r), 
\end{eqnarray} 
and the wave function $\psi$ representing the solution of the eigenvalue problem is separable in 
angular ($\theta$, $\phi$) and radial ($r$) coordinates:  $\psi = R(r) Y_{lm} (\theta, \phi$). Here, $R(r)$ 
  and $Y_{lm} (\theta, \phi)$ are radial and angular [given by spherical harmonics] wave functions, 
respectively. Note that the latter is the eigenfunction of the angular momentum operator $l^2$:
\begin{eqnarray}
l^2 Y_{lm} (\theta, \phi) = \hbar^2 l(l+1) Y_{lm} (\theta, \phi).
\end{eqnarray}  
The solutions of the hamiltonian with only central potential included are shown in the left column 
of Fig.\ \ref{s-p-spher}. The spectrum of  single-particle states, characterized by principal quantum 
number $N$, is equidistant in energy with high degree of the degeneracy of the single-particle states.

The realistic nuclear potential is located somewhat between pure HO and square 
well potentials. To correct for that  a centrifugal potential has to be added to $V(r)$;  it is usually 
parametrized as \cite{NilRag-book,N.55,RNS.78}:
\begin{eqnarray}
V_{corr} =  - \mu' \hbar \omega_{\,0} \left(l^2 - \left<l^2\right>_N\right).
\label{V-corr-1} 
\end{eqnarray}  
Within the $N$ shell this term leads to a lowering in energy of high-$l$ states relatively to low-$l$ ones
and to a  removal of the degeneracy between the $l$ states (see middle panel of Fig.\ \ref{s-p-spher}).

  In addition, there is a coupling between spin ($s=1/2$) and orbital motion of the single particle
which, in general,  is given by the following interaction term:
\begin{eqnarray}
V_{LS} = \lambda \frac{1}{r} \frac{\partial V_{SO}(r)}{\partial r} \,\, \vec{\it l} \cdot \vec{\it s}.
\label{spin-orbit}
\end{eqnarray}
Here $V_{SO}$ indicates the spin-orbit potential which may be different from the central
one and $\lambda$ is the coupling constant of spin-orbit interaction. In the case of harmonic oscillator 
potential this term can be further simplified to $V_{LS}= - 2\kappa \hbar \omega_{\,0} \vec{l} \cdot \vec{s}$.
 The presence of spin-orbit potential leads to the lowering and rising  in energy of the $j=l+1/2$ and $j=l-1/2$
states emerging from the state with a given orbital angular momentum $l$ (see right column in Fig.\ 
\ref{s-p-spher}). Only with this interaction included, it is possible to reproduce experimentally
observed shell closures at particle numbers 2, 8, 20, 28, 50, 82 and 126 \cite{Bohr1975,Casten-book,RNS.78}.
Note that the single-particle state is completely defined by a set of quantum numbers $[Nlj]$.

   The combined potential 
\begin{eqnarray}
V_{MO} = V(r) + V_{corr} + V_{LS} 
\end{eqnarray} 
is usually called as modified oscillator (MO) or Nilsson potential.  This potential is defined by three parameters
$\omega_0^i$, $\kappa_i$ and $\mu'_i$ for each kind of nucleons ($i=\pi$ or $\nu$). $\omega_0^i$ defines
the radius of respective matter distribution, $\mu'_i$ simulates the surface diffuseness depth, and $\kappa_i$
is the strength of spin-orbit interaction.  The Coulomb potential is not directly included into the MO
potential but it is effectively accounted by the differences of the above mentioned parameters in the 
proton and neutron subsystems.

   Although the MO potential is still extensively used in nuclear structure studies, more realistic 
treatment  of the single-particle degrees of freedom is achieved by means of the Woods-Saxon (WS) 
\cite{GISF.73,CDNSW.87,BDNO.89} 
and folded Yukawa (FY) \cite{MN.81,DPB.16} potentials. This is because they have  more 
realistic shape of the potential (thus eliminating the need for $V_{corr}$)
and explicitly include the Coulomb interaction.  Their structure is illustrated below 
for the WS potential 
\begin{eqnarray}
V_{WS} = V_{WS}(r) + V_{LS} + V_{C}, 
\end{eqnarray} 
where central potential is given by 
\begin{eqnarray}
V_{WS}(r) = \frac{V_0}{1+exp[(r-R)/a]}
\end{eqnarray}
with $V_0$ being the potential depth, $a$ the surface thickness and $R=r_0\,A^{1/3}$ 
the nuclear radius (with typical value of $r_0 \approx 1.2$ fm). Here $V_C$ 
represents the Coulomb potential. However, numerical realization of these potentials is
more complicated as compared with the MO potential. The detailed comparison of the 
WS and MO potentials is presented in Ref.\ \cite{BDNO.89}.

%%%%%%%%%%%%%%%%%%%%%%%%%
\section{Deformed shell model}
%%%%%%%%%%%%%%%%%%%%%%%%%

   In deformed nuclei,  the shape of the nuclear surface is generally 
parametrized by means  of a multipole expansion of the radius in 
terms of the shape parameters \cite{RS.80}; a typical parametrization is
\begin{eqnarray}
R(\theta, \phi)=R_0\left[ 1+\sum_{\lambda \mu} \alpha_{\lambda \mu}
Y_{\lambda \mu}(\theta, \phi) \right], 
\label{surf}
\end{eqnarray}
where $\alpha_{\lambda \mu}$ are the deformation parameters and $R_0$ is 
the radius of the sphere with the same volume.  For axially symmetric 
nuclear shapes $\mu=0$ and quadrupole deformation $\beta_2 = 
\alpha_{20}$ is dominant \cite{RS.80,CAFE.77} . For simplicity, higher multipolarity 
deformations such as octupole $\beta_3$, hexadecapole $\beta_4$, ... are 
not taken into account in the present discussion. Note that in the literature different 
parametrizations of the deformations of the single-particle potential exist (see 
discussion in Refs.\ \cite{NilRag-book,CAFE.77}).

   There are two facts which affect the consideration of single-particle potentials
in deformed nuclei \cite{N.55}. First, this potential should follow the nuclear density 
distribution. Second, in deformed nuclei the oscillator frequencies $\omega_{\,i}$ 
($i=x, y$ and $z$)  are different along different principal axis of nuclear ellipsoid.
As a result, the central potential of Eq. (\ref{central}) is modified in the 
following way:
\begin{eqnarray}
V(\vec{r}) = \frac{m}{2} \left(\omega_{\,x}^2 x^{\,2} + \omega_{\,y}^2 y^{\,2} + \omega_{\,z}^2 z^{\,2} \right).
\end{eqnarray}
Since nuclear matter is highly incompressible, the change of the shape of the
nucleus from spherical to ellipsoidal should not modify the volume of the nucleus.
This is accounted by the volume-conservation condition
\begin{eqnarray}
\omega_{\,0}^3 = \omega_{\,x} \omega_{\,y} \omega_{\,z}.
\end{eqnarray}
Assuming axial symmetry around the $z$ axis, i.e. $\omega_{\,x} = \omega_{\,y}$,  the
central potential can be rewritten as 
\begin{eqnarray}
V(\vec{r}) = \frac{1}{2} m\omega_{\,0}^2 r^{\,2} - \beta_2 m \omega_{\,0}^2 r^{\,2}  Y_{20} (\theta, \phi),
\end{eqnarray}
where $\beta_2$ stands for the quadrupole deformation of the potential
(the measure of the deviation from spherical shape).
Thus, the hamiltonian of the Nilsson model becomes
\begin{eqnarray}
h(\vec{r}) = - \frac{\hbar^2}{2m} \nabla_i^2 + \frac{1}{2} m\omega_{\,0}^2 r^{\,2} &-& 
\beta_2 m \omega_{\,0}^2 r^{\,2}  Y_{20} (\theta, \phi) \nonumber \\
&-& \mu' \hbar \omega_{\,0} \left(l^2 - \left< l^2\right>_N \right)
- 2\kappa \hbar \omega_{\,0} \vec{l} \cdot \vec{s}.
\end{eqnarray}
%
%Note that for simplicity I restricted myself to a general outline of the Nilsson model \cite{N.55}. 
Note that for simplicity only general outline of the Nilsson model is provided here
\cite{N.55}.  
%Thus,  higher multipolarity deformations such as octupole $\beta_3$, hexadecapole $\beta_4$, 
%... are not taken into account. 
Technical details of the solution of the Nilsson potential 
are discussed in Refs.\ \cite{NilRag-book,N.55,BDNO.89}.  Note also that 
there are generalizations of the Woods-Saxon and folded Yukawa potentials to deformed 
shapes \cite{GISF.73,CDNSW.87,MN.81,DPB.16,CAFE.77}.

\begin{figure}[t!]
\centering
\includegraphics[width=5.8cm]{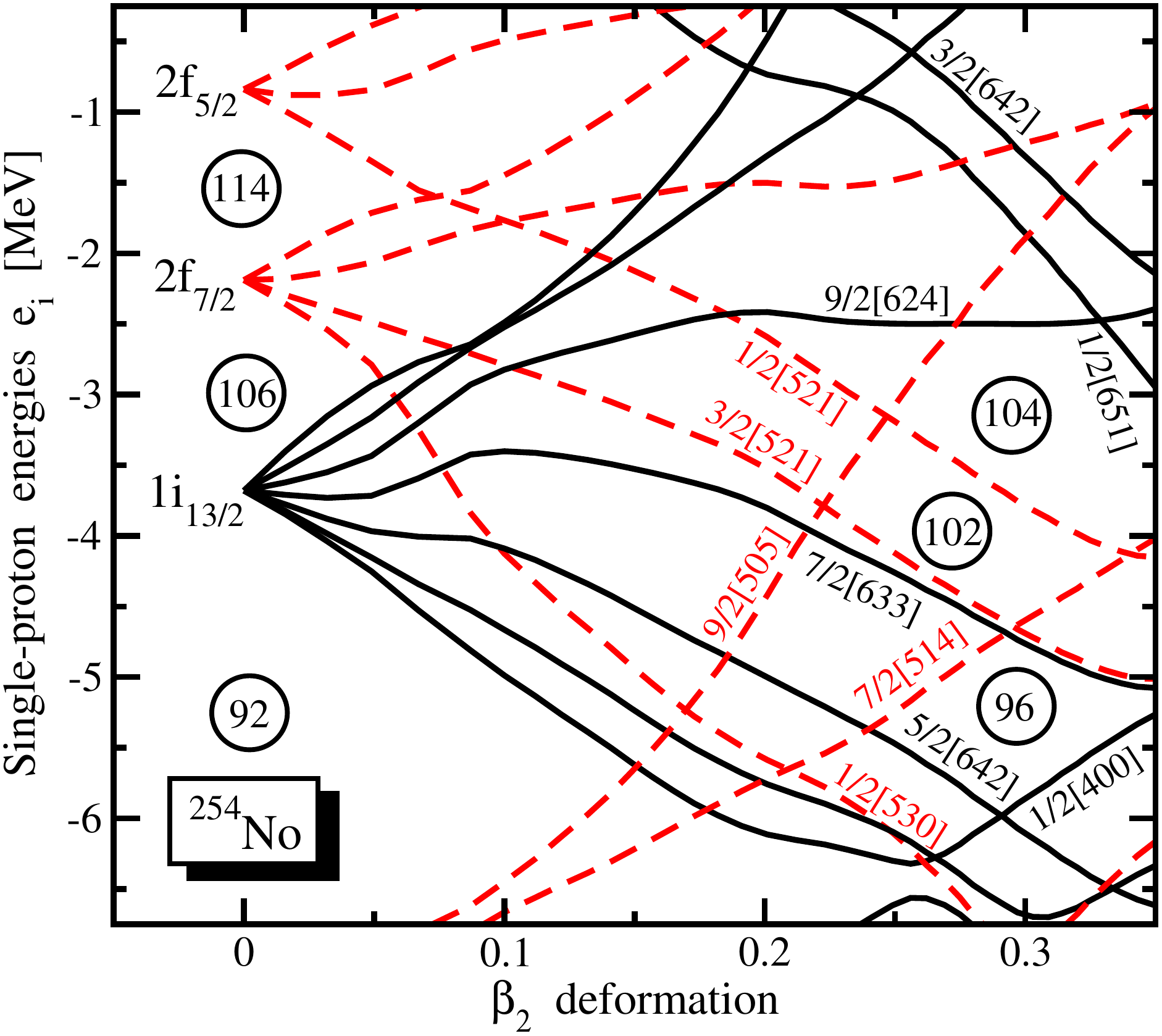}
\includegraphics[width=5.8cm]{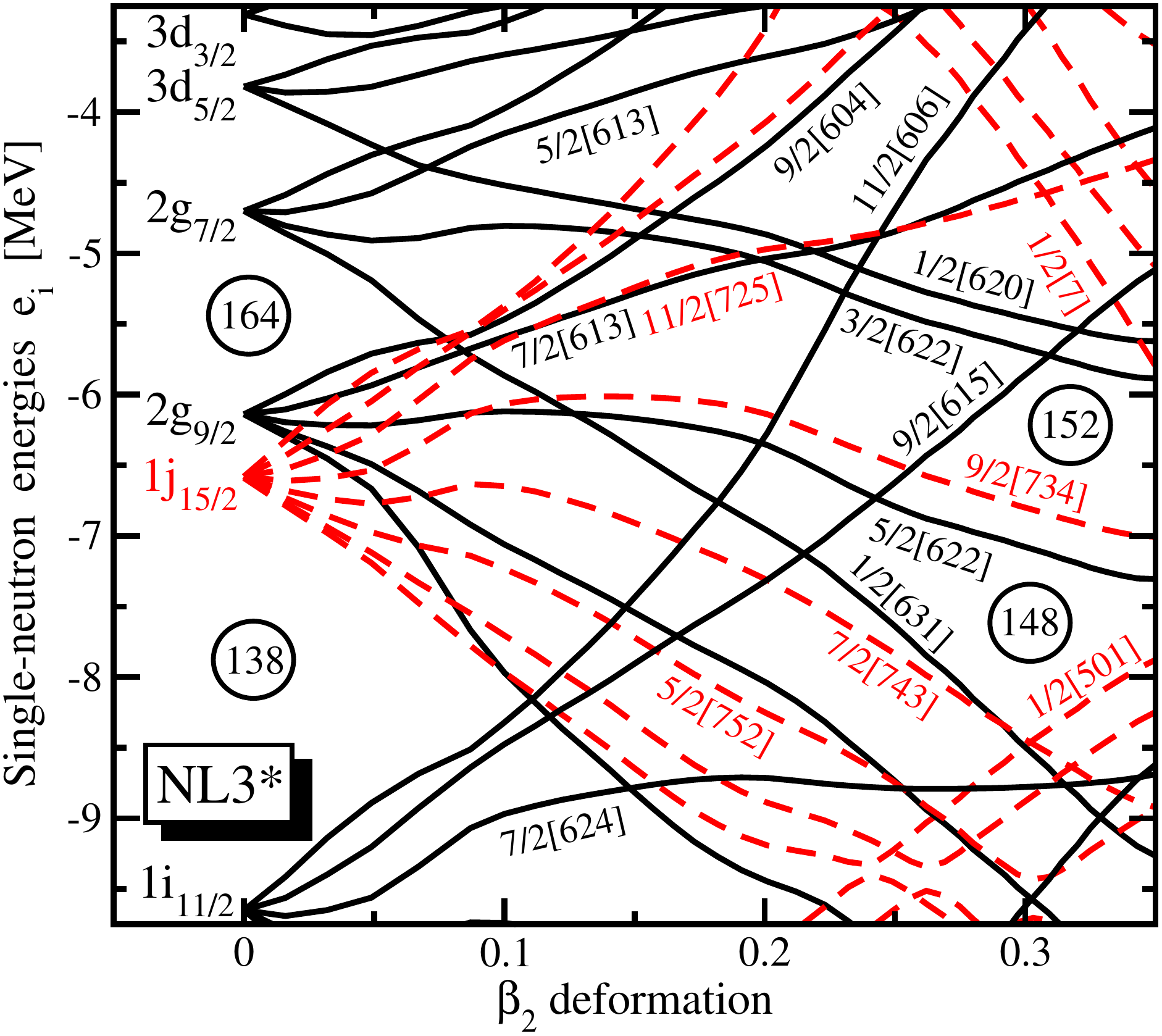}
\caption{Single-particle energies, i.e., the diagonal elements of the single-particle 
Hamiltonian $h$ in the canonical basis \cite{RS.80}, for the lowest in total energy 
solution in the nucleus $^{254}$No calculated as a function of the quadrupole equilibrium
deformation $\beta_2$ for covariant energy density functional NL3*. Solid and dashed 
lines are used for  positive- and negative-parity states, respectively. Relevant spherical 
and deformed gaps are indicated. Figure taken from Ref.\ \cite{DABRS.15}.
\label{s-p-def-254No}
}
\end{figure}

   As a result, the focus here is on the consequences of the breaking of 
spherical  symmetry on the single-particle states. They are usually illustrated by 
means of the Nilsson diagrams which show the evolution of the energies of deformed 
single-particle states as a function of deformation (see, for example, Figs. 3 and
5 in Ref.\ \cite{N.55}, Figs. 8.3 and 8.5 in Ref.\ \cite{NilRag-book} and Fig. 3
in Ref.\ \cite{PhysRep-SBT} for such diagrams obtained with the Nilsson potential).
The general structure (and frequently fine details) of the Nilsson diagrams  obtained 
in phenomenological  single-particle potentials and in self-consistent models 
are very similar.  This clearly indicates that the former have deep microscopic roots.
Thus, the results obtained in covariant density functional 
theory presented in Fig.\ \ref{s-p-def-254No} are used here to illustrate the impact of deformation
on the single-particle states.

%%%%%%%%%%%%%%%%%%%%%%%%%%%%%%%%%%%%%%%%% 
%%%%%%%%%%%%%%%%%%%%%%%%%%%%%%%%%%%%%%%%% 

   Several major features emerge on transition from spherical to deformed 
shapes.  First, it removes the $2j + 1$ degeneracy of the spherical subshells and deformed  
single-particle states are only two-fold degenerate. Second, the deformed single-particle 
states are defined by approximate Nilsson quantum numbers  $\Omega [Nn_z\Lambda]$
(note they are frequently shown in inverted order of $[Nn_z\Lambda]\Omega$).
Here, $\Omega$ ($\Lambda$) stands for the projection of the total (orbital) single-particle 
angular momentum on the axis of symmetry, $N$ is the principal quantum number  and $n_z$
is the number of nodes of the wavefunction along the symmetry axis. Note that the parity
$\pi$ of the state is defined by $(-1)^N$.  For prolate $(\beta_2>0)$ shapes, the deformed 
orbitals  emerging from a given spherical $j$-subshell are split in such a way that the state 
with $\Omega=1/2$ is always lowest in energy while that with $\Omega=j$ is the highest
one. The states with intermediate $\Omega'$ values are arranged in such a way
that the energies $E$ of two neighboring deformed states satisfy the condition $E(\Omega'+1) >
E(\Omega')$ (see Fig.\ \ref{s-p-def-254No}).The order of the states is inverted for the oblate
$(\beta_2 <0)$ shapes \cite{NilRag-book,N.55} . Third, as a consequence of these modifications of the 
single-particle states with deformation, spherical shell gaps disappear and new  deformed 
gaps, which are comparable in size with minor shell gaps at  spherical shape, appear.
These gaps are encircled in Fig.\ \ref{s-p-def-254No}).

%%%%%%%%%%%%%%%%%%%%%%%%%%%%%%%%%%%%%%%%%%%%%%%%%%%%%%%%%%
\subsection{Cranked shell model}
\label{crankingmodel}
%%%%%%%%%%%%%%%%%%%%%%%%%%%%%%%%%%%%%%%%%%%%%%%%%%%%%%%%%%

  Rotation is the phenomenon which appears in all branches of physics, from galaxies down
to atomic nuclei. In the latter, it is a collective phenomenon in which many nucleons define a nuclear
deformation and contribute to rotational motion.  Note that contrary to classical mechanics, a collective 
rotation along the symmetry axis of nuclear density distribution is forbidden in quantum mechanics. 

  The simplest way to describe the properties of rotating nuclei in a microscopic way is 
to use the cranking-model approximation suggested by Inglis \cite{Ing.54,Ing.56}. Over the years 
this model  has been successfully applied to the  description of different phenomena in rotating nuclei 
\cite{NilRag-book,PhysRep-SBT,VDS.83,Szybook,MPZZ.13}. The consideration here is restricted 
to  one-dimensional cranking in which the nuclear  field is rotated externally with a constant angular 
velocity $\omega$ around a principal axis usually defined as the $x$-axis. This is done in order to 
outline the basic  features of this model.

%%%%%%%%%%%%%%%%%%%%%%%%%%%%%%%%%%%%%%%%% 
\begin{figure}[t!]
\centering
\includegraphics[width=\linewidth]{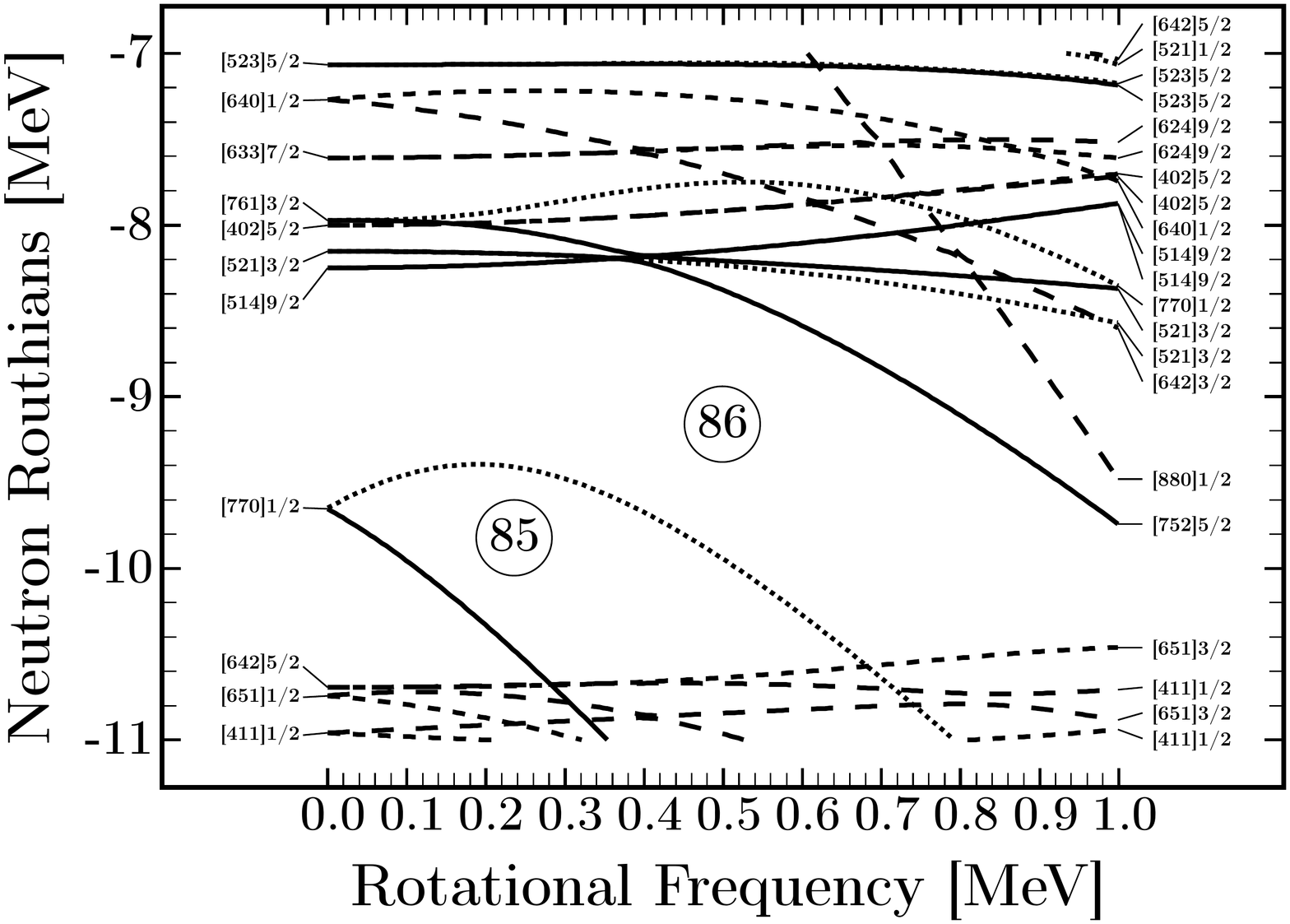}
\caption{Single-neutron energy levels obtained with the Woods-Saxon potential 
as a function of rotational frequency for deformation parameters $\beta_2=0.61$,
$\beta_4=0.11$ typical for  the yrast superdeformed band in $^{152}$Dy and its closest neighbors. 
The orbitals are defined by the parity $\pi$ and signature quantum number $\alpha$.  Solid, dotted, 
short-dashed and long-dashed lines are used for the orbitals with $(\pi=-,\alpha=-1/2)$
$(\pi=-,\alpha=+1/2)$, $(\pi=+,\alpha=+1/2)$ and $\pi=+,\alpha=-1/2)$, respectively. 
The dominant Nilsson components of the wavefunctions of the levels are shown at
the lowest and highest calculated frequencies. 
Figure is reproduced from Ref.\ \cite{151Tb}.
\label{s-p-rot}
}
\end{figure}
%%%%%%%%%%%%%%%%%%%%%%%%%%%%%%%%%%%%%%%%% 
 
    The basic idea of the cranking model is that a nucleus
with angular momentum $I\ne 0$ can be described in 
terms of an intrinsic state $\Psi^{\omega}$ at rest in a rotating 
frame. In the one-dimensional cranking approximation for collective rotation,
the total cranking Hamiltonian (or Routhian) for a  system of independent 
particles is given by
\begin{eqnarray}
H^{\omega} = H - \omega I_x = \sum_{\it i\,\,occ} h^{\omega}_i\,\,,
\label{TCH}
\end{eqnarray}
where $H$ is the total Hamiltonian in the laboratory system, $I_x$ 
is the $x$-component of the total angular momentum and 
$h^{\omega}$ is the single-particle Hamiltonian in the
rotating system
\begin{eqnarray}
h^{\omega}=h- \omega j_x
\label{SPCH}
\end{eqnarray}
with $h$ being the single-particle Hamiltonian in the laboratory 
system and $j_x$ the $x$-component of the single-particle angular 
momentum. In Eq.\ (\ref{TCH}), the sum extends over the occupied 
proton and neutron orbitals. The term $ -\omega I_x$ in 
Eq.\ (\ref{TCH}) is analogous to the Coriolis and centrifugal forces 
in classical mechanics.

Then the total energy $E_{tot}$ in the laboratory system is  given as
\begin{eqnarray}
E_{tot}=
\sum_{i\,\, occ} \langle \psi_i^{\omega} | h | \psi_i^{\omega} \rangle=
\sum_{i\,\, occ} e^{\omega}_i + \omega
\sum_{i\,\, occ} \langle \psi_i^{\omega} | j_x | \psi_i^{\omega} \rangle,
\label{ETOT}
\end{eqnarray}
and the total spin $I$ by
\begin{eqnarray}
I\approx I_x=\sum_{i\,\,occ}\langle 
       \psi_i^{\omega} | j_x | \psi_i^{\omega} \rangle,
\label{ITOT}
\end{eqnarray}
where $\psi_i^{\omega}$ are the single-particle eigenfunctions 
in the rotating system and $e^{\omega}_i=
\langle \psi_i^{\omega} | h^{\omega} | \psi_i^{\omega} \rangle$ 
the corresponding eigenvalues (single-particle routhians). The 
approximation used in Eq.\ (\ref{ITOT}) is valid only in the limit 
of high spin ($I\gg 1$).

The evolution of the single-particle states in rotating potential is
shown in Fig.\ \ref{s-p-rot}.  There are several important features. 
First, the time-reversal symmetry is broken in rotating potential and
two-fold degeneracy of deformed single-particle states, which exist
in non-rotating nuclei, is removed. As a consequence, each single-particle  
orbital has additionally to be characterized by signature quantum 
number (either $r_i$ or $\alpha_i$) \cite{NilRag-book,PhysRep-SBT,VDS.83}. 
This quantum number is a consequence of the fact that full cranking 
Hamiltonian is invariant with respect to a rotation through an angle $\pi$ 
around the cranking axis ($x$-axis)
\begin{eqnarray}
{\cal R}_x=\exp (-i \pi j_x),\,\,\,\,\,\, 
{\cal R}_x \psi_i=\exp (-i \pi j_x) \psi_i\,\,.
\label{Rx}
\end{eqnarray}
The eigenvalues of ${\cal R}_x$ are $\exp (-i \pi \alpha)$, where $\alpha$ is the 
signature exponent quantum number. Alternatively, one can define signature 
quantum number as $r=\exp (-i \pi \alpha)$. Signature quantum number 
$\alpha_i$ ($r_i$) of a single-particle orbital could take the $\alpha_i=+\frac{1}{2}$
$(r_i=-i)$ or $\alpha_i=-\frac{1}{2}$ $(r_i=+i)$ values.  Similar to parity, this
classification of the single-particle states is an important tool for identifying 
the nucleon orbitals in the rotating nuclear potential.

  Second, the coupling between the different single-particle orbitals increases 
with increasing rotational frequency.  As illustrated in Fig.\  \ref{s-p-rot} this leads 
to the change of the dominant Nilsson components of the wavefunction.  
Third,  the slope of the orbitals in Fig.\ \ref{s-p-rot}  corresponds to the single-particle
alignment $\left< j_x \right>_i$ \cite{NilRag-book,VDS.83} 
\begin{eqnarray}
\left< j_x \right>_i = - \frac{\partial e_i^{\omega}}{\partial \omega} 
\label{slope} 
\end{eqnarray}   
for the case of cranking at fixed deformation. Note that single-particle routhians
are always plotted at fixed deformation in phenomenological potentials. In 
contrast, they are given along the equilibrium deformation path in the DFT
calculations.  Thus, the condition (\ref{slope}) is only approximately satisfied 
in the latter case provided that  deformation changes with increasing
rotational frequency are modest. Note  that in general  $\left< j_x \right>_i$ 
depends on signature and this dependence is especially pronounced for 
the single-particle orbitals with low value of $\Omega$.

   The knowledge of single-particle alignments is extremely useful for an understanding 
of physical situation and interpretation of experimental data. For example, high-$j$ or 
high-$N$ intruder orbitals (such as those with Nilsson labels $1/2[770]$ and $1/2[880]$) 
have large values of  $\left< j_x \right>_i$ and, as a consequence, they are strongly
downsloping with increasing rotational frequency (see Fig.\ \ref{s-p-rot}). Large
energy splitting between two signature partner orbitals of a given single-particle
state appear for the states with $\Omega=1/2$ but no such splitting exist
for the states with large value of $\Omega$. For example, this is a case for 
the $1/2[770]$ state for which a large signature splitting between the $\alpha=\pm1/2$ 
branches leads to a formation of large $N=85$ gap at $\omega \approx 0.3$ MeV which is 
absent at $\omega=0$ MeV (see Fig.\ \ref{s-p-rot}).

   The advantage of the cranking model is that it provides 
a microscopic description of rotating  nuclei, where  the total angular momentum is 
described as a sum of single-particle angular momenta, and thus collective and 
`non-collective' rotations can be treated on the same footing. On the other hand, 
there are limitations. The cranking-model approximation is semiclassical, because 
the rotation is imposed externally. The model also breaks the rotational invariance, 
since a fixed rotation axis is used. These latter limitations are not very important for 
very fast rotation $(I\gg 1)$. Another shortcoming of the cranking model is that the 
wave functions are not  eigenstates of the angular momentum operator, which can 
lead to difficulties; for example, a proper calculation of electromagnetic transition
probabilities requires the use of projection techniques~\cite{RS.80}. Furthermore, 
the cranking model is a poor approximation in the crossing region of two weakly 
interacting bands, but this difficulty can be overcome by the removal of such crossings 
\cite{PhysRep-SBT}.

%%%%%%%%%%%%%%%%%%%%%%%%%%%%%%%%%%%%%%%
\subsection{Spatial densities of the single-particle states}
\label{s-p-states}
%%%%%%%%%%%%%%%%%%%%%%%%%%%%%%%%%%%%%%%

   The total density of the nucleus is defined as the sum of the single-particle
densities of occupied states
\begin{eqnarray}
\rho_{tot}(\vec{r}) = \sum_{k} \rho_k(\vec{r}),
\end{eqnarray} 
which are given by
\begin{eqnarray}
\rho_k(\vec{r}) = \psi_k^*(\vec{r})  \psi_k(\vec{r})
\end{eqnarray} 
for the $k-$th single-particle state.
The single-particle wave function  $\psi_k(\vec{r})$ can be expanded into
basis states $|\mu>$
\begin{eqnarray}
\psi_k(\vec{r}) = \sum_{\mu} c_{k\mu} |\mu>,
\label{expansion}
\end{eqnarray} 
where $\mu$ represents the set of quantum numbers defining the basis 
state and $c_{k\mu}$ the expansion coefficients. Of special interest are the 
cases when this expansion is dominated by a single basis state $n$ ($c_{kn} \approx 1$)
since then the nodal structure of the wave function  of the state $\psi_k(\vec{r})$  and 
consequently its single-particle density will be predominantly defined by this basis state. 
This takes place  either at spherical shape or at extremely elongated nuclear shapes 
typical for rod shape structures or megadeformation (MD) \cite{AA.18}.

%%%%%%%%%%%%%%%%%%%%%%%%%%%%%%%%%%%%%%%%% 
\begin{figure}[t!]
\centering
\includegraphics[width=\linewidth]{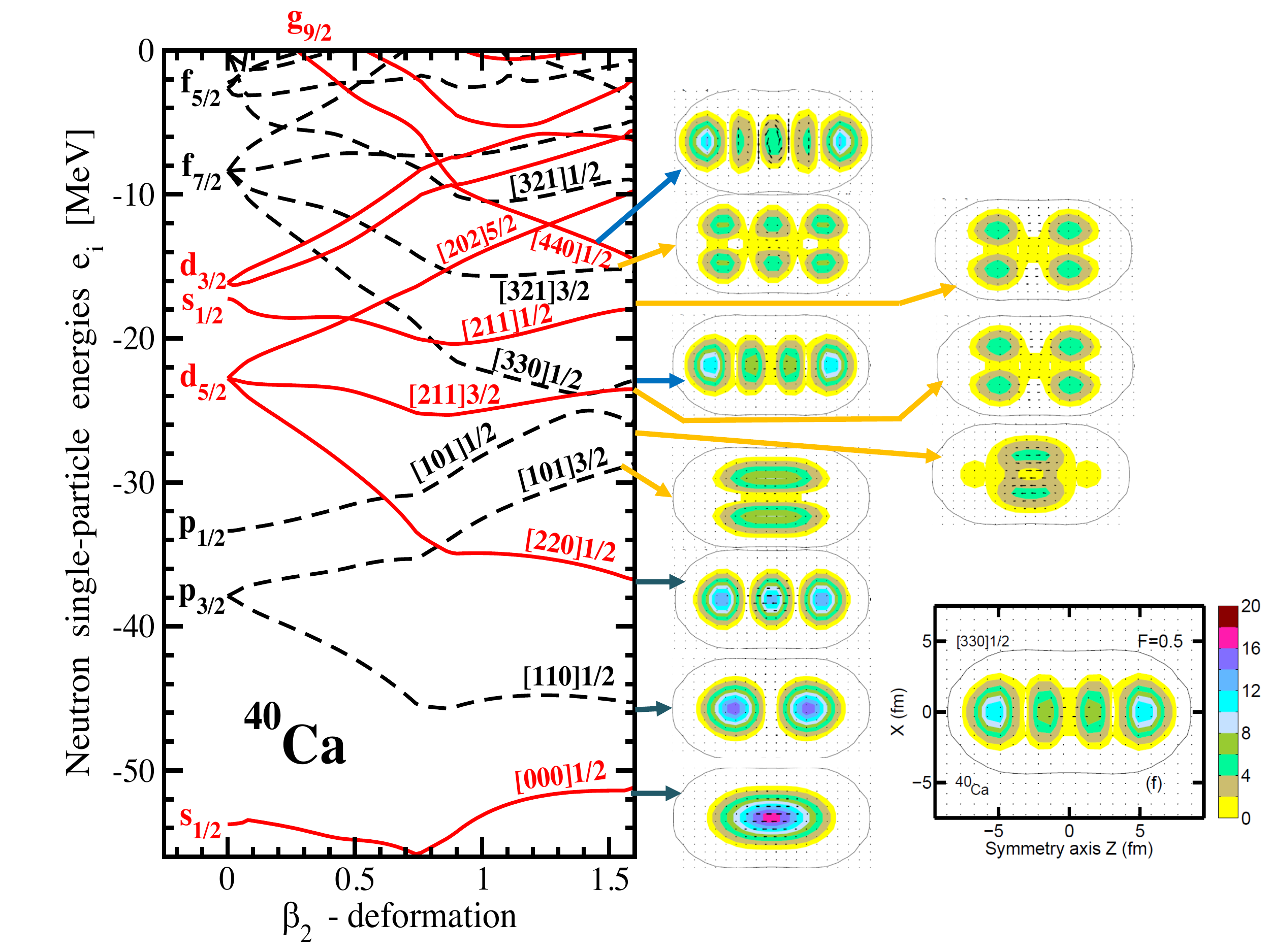}
\caption{
Left panel: The Nilsson diagram for the neutron single-particle states in 
$^{40}$Ca. It  is based on the results obtained in axial relativistic Hartree-Bogoliubov calculations 
with the NL3* functional. Right panel: Single-neutron density distributions due to the occupation 
of indicated  Nilsson states.The box in right bottom corner  exemplifies physical dimensions of 
the  nucleus as well as the colormap used for single-particle densities. Other density plots are 
reduced down to the shape and  size of the nucleus which is indicated by black  solid line 
corresponding to total neutron density line of $\rho=0.001$ fm$^{-3}$. The colormap shows
the densities as multiplies  of  $0.001$ fm$^{-3}$; the plotting of the densities starts with yellow 
color at $\rho=0.001$ fm$^{-3}$. Figure taken from Ref.\  \cite{A.18} and it is based 
on the results presented in Ref.\ \protect\cite{AA.18}.
\label{nodal}
}
\end{figure}
%%%%%%%%%%%%%%%%%%%%%%%%%%%%%%%%%%%%%%%%% 

    In the latter case, the wave function defined by asymptotic Nilsson 
quantum  number is expanded into the basis states characterized by 
$\mu =[Nn_z\Lambda]\Omega$ \cite{NilRag-book,AA.18}. For extremely 
elongated shapes in light nuclei this expansion is dominated by a single 
basis state \cite{AA.18}.  This is illustrated in Fig.\ \ref{nodal} which shows the
single-particle densities of indicated states at the deformation $\beta_2 \approx 1.6$ 
typical for the MD shapes. They are characterized by  axial or nearly axially symmetric 
spheroidal/ellipsoidal like density clusters formed for the single-particle states with 
the $[NN0]1/2$ Nilsson quantum numbers, doughnut density distributions for the
$[N01]\Omega$ states, multiply (two for $n_z=1$ and three for $n_z=2$) 
ring shapes for the $[N,N-1,1]\Omega$ Nilsson states with $N=2$ and 3. Fig.\ \ref{nodal} 
clearly indicates the importance of the deformation which has two critical effects. First, it 
leads to the formation of density clusters with specific nodal structure and to the 
separation of the clusters in space. Second, it lowers the energies of the Nilsson states 
of the$[NN0]1/2$ type which favors the $\alpha$-clusterization and leads to the 
configurations in which all occupied single-particle states have this type of structure.

 %%%%%%%%%%%%%%%%%%%%%%%%%%%%%%%%%%%%%%%%% 
\begin{figure}[t!]
\centering
\includegraphics[width=5.0cm]{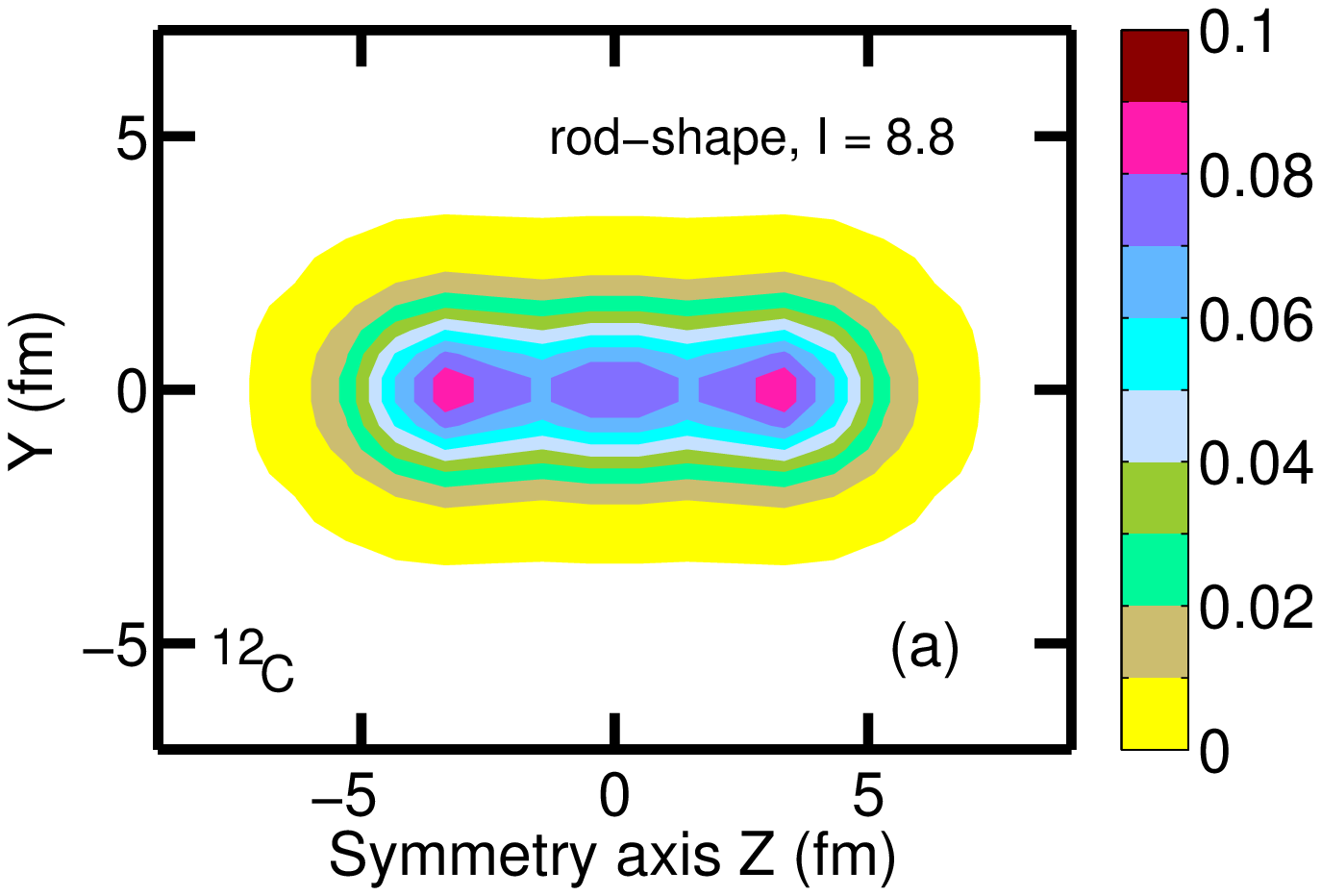}
\includegraphics[width=5.0cm]{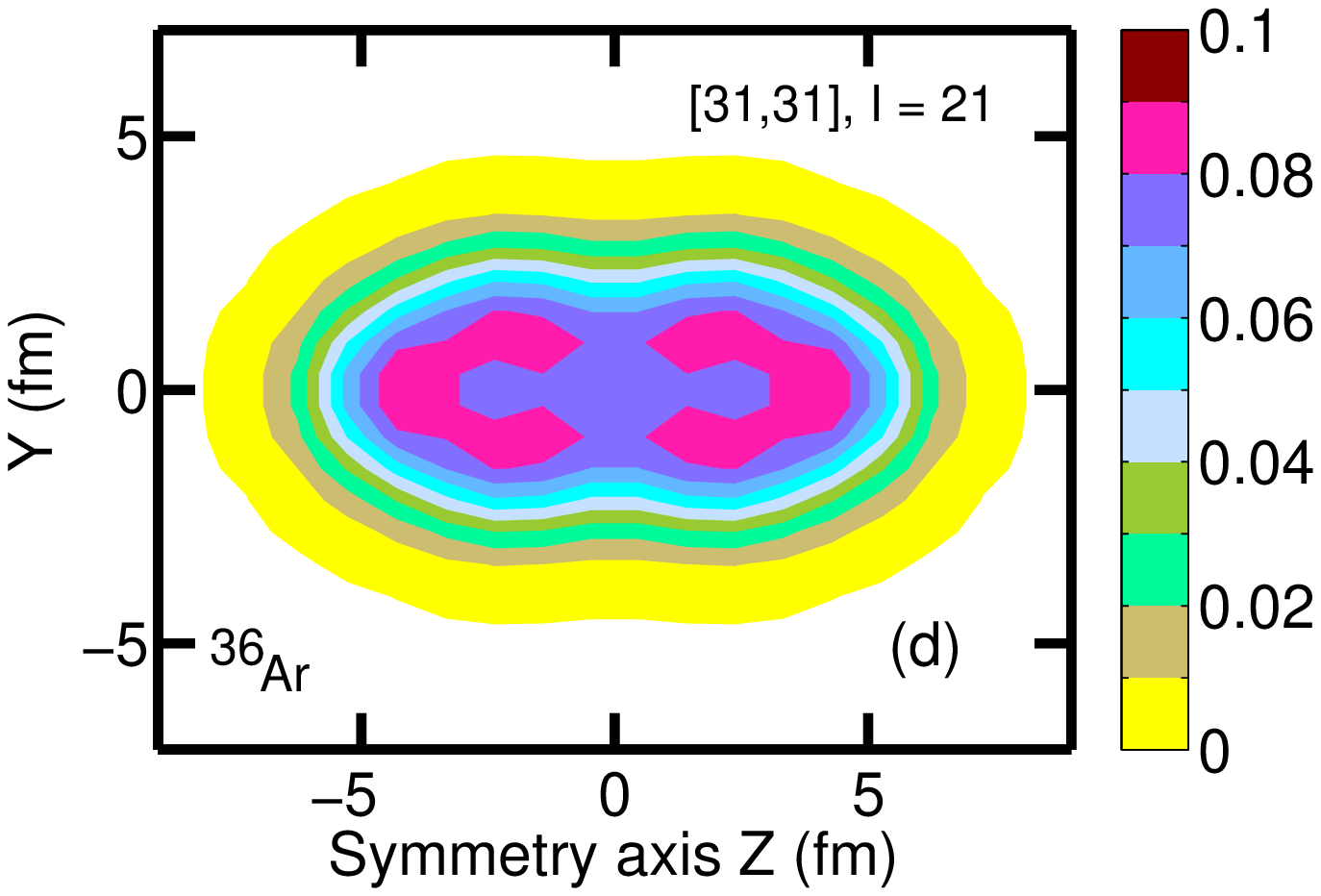}
\caption{Total neutron densities  [in fm$^{-3}$] of indicated configurations
in the $^{12}$C and $^{36}$Ar nuclei at specified spin values. The plotting
of the densities starts with yellow color at 0.001 fm$^{-3}$. The results are
based on the cranked relativistic mean field calculations with the NL3* functional. 
Figure taken from Ref.\  \protect\cite{AA.18}. 
\label{density-light}
}
\end{figure}
%%%%%%%%%%%%%%%%%%%%%%%%%%%%%%%%%%%%%%%%% 

%%%%%%%%%%%%%%%%%%%%%%%%%%%%%%%%%%%%%%%%% 
\begin{figure}[t!]
\centering
\includegraphics[width=6.0cm]{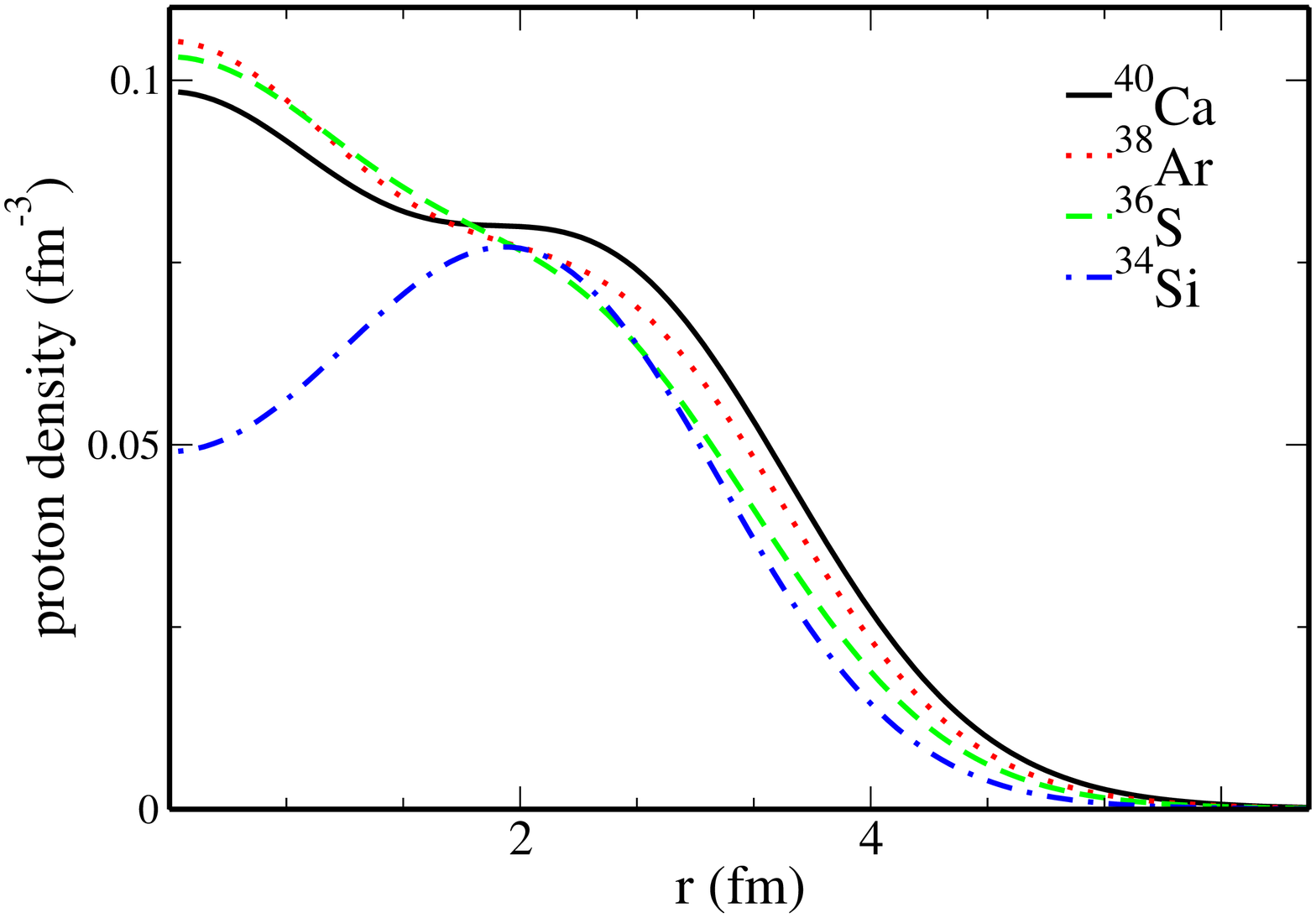}
\includegraphics[width=5.0cm]{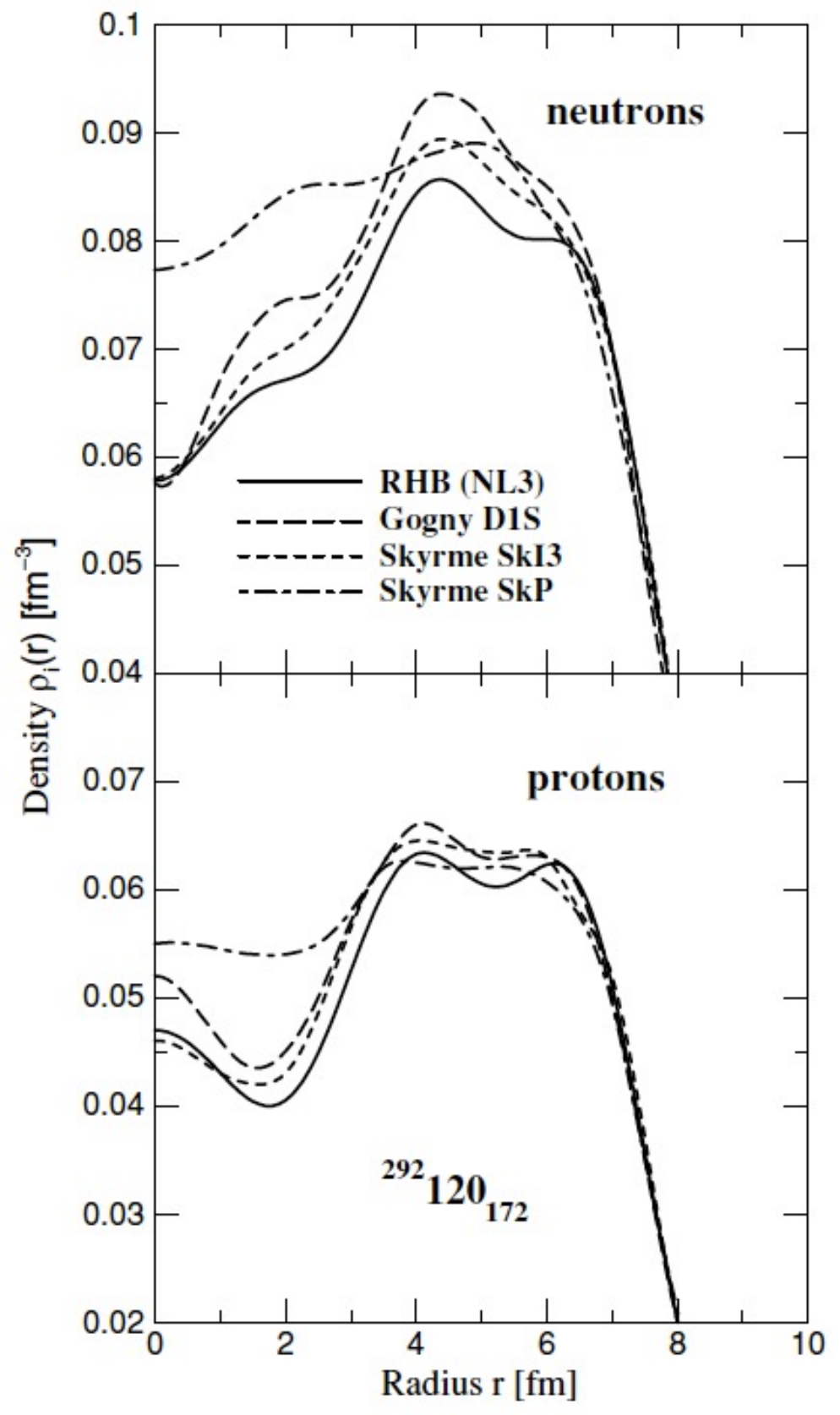}
\caption{Left panel: Proton densities of the $N=20$ isotones obtained in the spherical 
relativistic Hartree-Bogoliubov  calculations with the DD-ME2 functional. Right panels: 
Neutron  and proton densities of the spherical $^{292}120_{172}$ nucleus obtained in 
non-relativistic Skyrme and Gogny as well as covariant DFT calculations. Employed 
functionals are indicated. Figures are taken from Refs.\  
\protect\cite{KLRL.17,AF.05-dep}. 
\label{density-bubble}
}
\end{figure}
%%%%%%%%%%%%%%%%%%%%%%%%%%%%%%%%%%%%%%%%% 

It is impossible to experimentally measure the single-particle density 
distributions in rotating nuclei. That is a reason why contrary to the single-particle energies 
and alignments  they  have not been used as fingerprints of independent particle motion. However, 
they are important for an understanding of the $\alpha$-clusterization and the evolution of 
extremely elongated shapes in light nuclei such as ellipsoidal and rod shapes and nuclear 
molecules \cite{AA.18,A.18}. The examples of total neutron densities of later two types of 
nuclear shapes are provided in Fig.\ \ref{density-light}.  A rod-shaped nuclear configuration 
in $^{12}$C is built from a linear chain of three $\alpha$-clusters and could be well understood 
from the summation of the single-particle densities of occupied single-particle states \cite{AA.18} 
(see also Refs.\  \cite{IMIO.11,YIM.14,ZIM.15} for additional discussion of rod-shaped nuclei 
built on linear  chains of the $\alpha$ particles). Rod-shaped  configurations appear also in heavier 
nuclei such as $^{42}$Ca and $^{44}$Ti \cite{AA.18}. However, their densities are more uniform 
since the $\alpha$-clusterization  is suppressed by the occupation of the single-particle orbitals 
which have either doughnut or ring type single-particle density distributions (see Fig.\ \ref{nodal}
and Refs.\ \cite{AA.18,EKLV.18}).  Nuclear  molecules are  characterized by the formation of the 
neck between two nuclear fragments:  the configuration  of $^{36}$Ar is an example of such structure 
(see Fig.\ \ref{density-light}). 

     A knowledge of the nodal structure of the single-particle densities of the states allows
to understand the microscopic mechanism of the transition between different nuclear shapes. 
For example,  in order to build nuclear molecules from typical ellipsoidal density distributions 
one has to move matter from the neck (equatorial region) into the polar one. This can be achieved 
by means of specific particle-hole excitations which move the particles from (preferentially) doughnut 
type orbitals or from  the orbitals which have a density ring in an equatorial plane into the orbitals 
(preferentially of the $[NN0]1/2$ type) building the density mostly in the polar regions of the nucleus 
\cite{AA.18,A.18}.

    Another example of important role of the single-particle density of the states is seen in 
so-called ``bubble" structures of the ground states of spherical nuclei. This is due to a specific 
radial nodal structure of the wavefunctions of the single-particle states (see Fig. 6.2 in Ref.\ 
\cite{NilRag-book}): only $s$ ($l=0$) states build the density in the center of the nucleus. On 
the contrary, the $l>0$ states do not participate in the building of the densities in the center 
of the nucleus due to presence of the  centrifugal barrier (see also Fig. 2 in Ref.\ \cite{KLRL.17}).  
The impact of this feature is especially pronounced in proton subsystem of $^{34}$Si 
\cite{KLRL.17,GGKNPOGV.09}. Left panel of Fig.\ \ref{density-bubble}  shows the evolution of 
the proton densities in the $N=20$  isotones. The $^{40}$Ca, $^{38}$Ar and $^{36}$S  nuclei 
in which the proton $2s_{1/2}$ orbital is occupied show similar proton density patterns with the density 
peak at the center. In contrast, the emptying of the proton $2s_{1/2}$  orbital in $^{34}$Si  leads 
to a considerable depletion of the proton density in the central region of nucleus. This ``bubble" 
phenomenon has been indirectly confirmed in experiment \cite{Bubble-exp}.

   Similar depletion of the density in the central region of the nucleus exist also in superheavy 
nuclei (see Refs.\ \cite{AF.05-dep,BRRMG.99,SNR.17} and right panel of Fig.\ \ref{density-bubble}). 
However, the mechanism of its creation is different as compared with the one active
in $^{34}$Si since the emptying of the $s$ states is not possible in such  nuclei.   It relies on 
the fact that the filling of low-$j$/high-$j$ orbitals builds the density in the central/surface region of 
the nucleus \cite{AF.05-dep,BRRMG.99}). Thus, starting from the flat density distribution in $^{208}$Pb 
(which is experimentally verified, see Fig. 2.4 in Ref.\ \cite{NilRag-book}) one can build ``bubble"-type 
structure in the $^{292}120_{172}$ superheavy nucleus by occupying predominantly high-$j$ orbitals 
outside the $^{208}$Pb core \cite{AF.05-dep}.  It was suggested in Ref.\ \cite{SNR.17} that the 
depletion of density in central region is mostly due to Coulomb interaction. This, however, contradicts 
to the observation that spherical superheavy nuclei with $Z=126$ have significantly smaller depletion 
of the density in the central region as compared with the $^{292}120_{172}$ one \cite{AF.05-dep}.

%%%%%%%%%%%%%%%%%%%%%%%%%
\section{Microscopic+macroscopic models}
%%%%%%%%%%%%%%%%%%%%%%%%

   The "shell model" concepts discussed above assume some fixed mean
field properties which do not depend on nucleonic configuration. In addition,
they require the use of deformation parameters in the case of deformed and
cranked shell models which have to be defined either from experimental data
or from higher level model.

  To overcome these deficiencies the microscopic+macroscopic (mic+mac) 
model has been  suggested in 1960s \cite{Strut67,Nil69,Brack72} which 
can be considered as an approximation to the Hartree-Fock approach \cite{Brack72}.
In this model the total energy of the nucleus $E_{tot}$ is separated into two 
parts, a macroscopic part $E_{macro}$ and a microscopic part $E_{micro}$,
\begin{eqnarray}
E_{tot}=E_{macro}+E_{micro}.
\label{etot} 
\end{eqnarray}
The macroscopic energy $E_{macro}$ is defined by some version of the liquid-drop 
model while the microscopic energy $E_{micro}$ is obtained from  quantal shell 
corrections $E_{sh}$ calculated from a phenomenological  potential using 
the Strutinsky prescription \cite{Strut67}. The shape of the nuclear surface is
parametrized by means of a multipole expansion of the  radius in terms of the 
shape parameters (see Eq.\ (\ref{surf})). Then the  equilibrium deformation in a 
specific nucleonic configuration is  determined by a minimization of the total energy 
$E_{tot}$ with respect to the shape parameters.

  For a simplicity, only  the basic features of the mic-mac  method for the rotating 
nuclei are outlined here since the case of no rotation is easy to obtain by dropping respective  terms. 
The total nuclear energy $E_{tot}$ at a specific deformation 
$\bar{\beta}=\beta_2, \gamma, \beta_4, \ldots$ and spin $I_0$ is given  as a sum of 
the  rotating liquid drop energy and the shell energy 
\begin{eqnarray}
E_{tot} \left( \bar{\beta},I_0\right) = E_{LD} \left(
\bar{\beta},I=0\right) + \frac{1}{2{\cal
J}_{rig}(\bar{\beta})}I_0^2+ 
E_{sh} \left( \bar{\beta},I_0 \right).
\label{ETOTmacrmicr}
\end{eqnarray}
The shell energy is defined as the difference between the
discrete and smoothed (indicated by\,\,\, $\widetilde{     
}$\,\, ) single-particle energy sums, 
\begin{eqnarray}
E_{sh} \left( I_0 \right) = \left. \sum e_i
(\omega, \bar{\beta})
\right|_{I=I_{0}} - 
\left. \widetilde{\sum e_i
(\widetilde{\omega}, \bar{\beta})
} \right|_{\widetilde{I}=I_{0}}
\label{Eshell}
\end{eqnarray}
with both terms evaluated at the same spin value $I_0$.
The smoothed sum is calculated using the Strutinsky procedure \cite{Strut67}. 
Then the total nuclear energy $E_{tot} \left( \bar{\beta},I_0\right)$ and
equilibrium deformation $\bar{\beta}_{eq}$ of specific
nucleonic configuration can be calculated as a function of spin $I$ (defined 
via Eq.\ (\ref{ITOT})). By considering numerous configurations,
defined by the occupation of single-particle orbitals with the specific sets of
quantum numbers such as parity, signature etc,  one can build comprehensive 
spectra of multiply rotational bands and compare them with experimental data.

   Note that in the mic+mac approach the moment of inertia, defined from smoothed 
single-particle quantities, has to be renormalized to a rotating liquid behavior. The renormalization 
of the moment of inertia is especially important in the Nilsson potential because of the $l^2$-term  
but it is also required in the Woods-Saxon potential because the potential radius is often different 
from the nuclear matter radius (see Ref.\ \cite{PhysRep-SBT} and references therein).

 The consideration above is restricted to the situation of no pairing correlations. The
experience and detailed studies show that this is quite accurate approximation at high
spins \cite{PhysRep-SBT,VALR.05}. However, the pairing correlations play an important 
role at low and medium spins. In the mic+mac approaches they are usually taken into 
account at the BCS level by adding pairing energy term $E_{pairing}$ 
\cite{PD.03,CRBLLP.08,MSIS.16} 
\begin{eqnarray}
E_{tot}^{pair}=E_{tot}+E_{pairing}.
\label{etot} 
\end{eqnarray}
The mic+mac model is simpler than self-consistent approaches and it is also substantially 
cheaper in numerical calculations. Despite approximations 
and simplifications employed by this method, it provides an accurate description of 
the ground state energies and deformations as well as fission barrier heights 
\cite{PD.03,MSIS.16} and rotational properties of multiply bands in different nuclei
\cite{PhysRep-SBT,CRBLLP.08}.
 
%%%%%%%%%%%%%%%%%%%%%%%%%%%%%%%%%%%%%
\section{Self-consistent approaches: covariant density functional theory}
%%%%%%%%%%%%%%%%%%%%%%%%%%%%%%%%%%%%%

   There are several types of self-consistent density functional theories (DFTs) 
based either  on the non-relativistic Shr\"{o}dinger equation with finite range Gogny 
or zero range  Skyrme forces  or on the relativistic Dirac equation. The latter is 
called as covariant density functional theory (CDFT) \cite{VALR.05} and its brief
outline will be presented here. It was very successful in the description of many
physical phenomena (see reviews in Refs.\ \cite{VALR.05,MTZZLG.06,NVR.11,RDFNS.16}). 
The detailed reviews of the Skyrme and Gogny DFTs are presented in Refs.\  
\cite{BHP.03,PM.14}.   

    There are several classes of the CDFT models and, for simplicity, only 
meson-exchange (ME) models will be considered here. In these models the nucleus 
is described as a system of Dirac nucleons interacting via the exchange of mesons 
with finite masses leading to finite-range interactions. The starting point of the ME 
models is a standard Lagrangian density \cite{GRT.90}
\begin{eqnarray}
\label{lagrangian}%
\mathcal{L}    = && \bar{\psi}\left[%
\gamma\cdot(i\partial-g_{\omega}\omega-g_{\rho
}\vec{\rho}\,\vec{\tau}
%-eA
-e \frac{1-\tau_3} {2} A
)-m-g_{\sigma}\sigma
\right] \psi
+\frac{1}{2}(\partial\sigma)^{2}-\frac{1}{2}m_{\sigma}^{2}\sigma^{2} \nonumber \\
&&-\frac{1}{4}\Omega_{\mu\nu}\Omega^{\mu\nu}+\frac{1}{2}m_{\omega}^{2}\omega^{2} 
-\frac{1}{4}{\vec{R}}_{\mu\nu}{\vec{R}}^{\mu\nu}+\frac{1}{2}m_{\rho}^{2}\vec{\rho}^{\,2}
-\frac{1}{4}F_{\mu\nu}F^{\mu\nu}, 
\end{eqnarray}
which contains nucleons described by the Dirac spinors $\psi$ with the mass $m$ and 
several effective mesons characterized by the quantum numbers of spin, parity, and 
isospin. The Lagrangian (\ref{lagrangian}) contains as parameters the meson masses $m_{\sigma}$, 
$m_{\omega}$, and $m_{\rho}$ and the coupling constants $g_{\sigma}$, 
$g_{\omega}$, and $g_{\rho}$. $e$ is the charge of the protons and it vanishes
for neutrons. The coupling constants are density dependent in the density dependent
meson exchange (DDME) class 
of covariant  energy density functionals (CEDFs) \cite{TW.99,DD-ME2}. In contrast,  they 
are constant in the so-called non-linear (NL) CEDFs in which the density dependence is 
introduced  via the powers of the $\sigma$-meson \cite{BB.77}:
\begin{eqnarray}
\mathcal{L} _{NL} =  \mathcal{L}  - \frac{1}{3} g_2 \sigma^3 - \frac{1}{4} g_3 \sigma^4.
\end{eqnarray}

The solution of these Lagrangians leads to the relativistic Hartree-Bogoliubov (RHB) 
equations \cite{VALR.05}. They are illustrated below on the example of the cranked RHB
(CRHB) equations for the fermions in the rotating  frame (in one-dimensional cranking 
approximation)   \cite{CRHB}
\begin{eqnarray}
\begin{pmatrix}
  \hat{h}_D-\lambda_{\tau}-\Omega_x \hat{J_x} & \hat{\Delta} \\
 -\hat{\Delta}^*& -\hat{h}_D^{\,*} +\lambda_{\tau}+\Omega_x \hat{J_x}^* 
\end{pmatrix} 
\begin{pmatrix} 
U_k({\bm r})\\ V_k({\bm r})
\end{pmatrix}
%\nonumber \\ 
= E_k
\begin{pmatrix} 
U_k({\bm r}) \\ V_k({\bm r}) 
\end{pmatrix},\,\,\,\,\,\,\,\,\,
\end{eqnarray}
where  $\lambda_{\tau}$ ($\tau=p,\,\,n$) are chemical potentials defined 
from the average particle number constraints for 
protons and neutrons; 
$U_k ({\bm r})$ and $V_k ({\bm r})$ are quasiparticle 
Dirac spinors; $E_k$ denotes the quasiparticle energies; and $\hat{J_x}$
is the angular momentum component entering into the Coriolis term
$-\Omega_x J_x$. Here,  $\hat{h}_D$ is the Dirac Hamiltonian for the 
nucleon with mass $m$
\begin{eqnarray}
  \hat{h}_D = \alpha (-i\nabla - \bm V(\bm r)) + V_0(\bm r) + \beta (m + S(\bm r)), 
\end{eqnarray}
which contains an attractive scalar potential 
$S({\bm r})$
\begin{eqnarray}
S(\bm r)=g_\sigma\sigma(\bm r),
\label{Spot}
\end{eqnarray}
a repulsive vector potential $V_0({\bm r})$
\begin{eqnarray}
V_0(\bm r)~=~g_\omega\omega_0(\bm r)+g_\rho\tau_3\rho_0(\bm r)
+e \frac{1-\tau_3} {2} A_0(\bm r),
\label{Vpot}
\end{eqnarray}
and a magnetic potential ${\bm V}({\bm r})$
\begin{eqnarray}
\bm V(\bm r)~=~g_\omega\bm\omega(\bm r)
+g_\rho\tau_3\bm\rho(\bm r)+
e\frac {1-\tau_3} {2} \bm A(\bm r).
\label{Vmag}
\end{eqnarray}

    These equations are solved numerically in triaxial harmonic oscillator basis 
and signature basis is used for the single-particle states. The CRHB framework is
applicable to the description of both rotating nuclei in the paired regime and 
one-/two-quasiparticle configurations in rotating and non-rotating nuclei. An approximate particle 
number projection by means of the Lipkin-Nogami (LN) method is also used in it 
\cite{CRHB} but its discussion is omitted for simplicity. The CRHB+LN calculations 
have been successful in the description of rotating nuclei in paired regime 
\cite{VALR.05,CRHB,AO.13} and one-quasiparticle configurations in odd-$A$ nuclei
(see discussion of Fig.\ \ref{s-p-def} below). Moreover, the CRHB framework can be either 
reduced to unpaired regime, leading to so-called cranked relativistic mean field (CRMF) approach
\cite{KR.89,ALR.98}, or upgraded to the 2- and 3-dimensional cranking approximation 
\cite{MPZZ.13}. By putting $\Omega_x=0$ it can also be applied to the description of 
non-rotating nuclei, but more specialized versions of the RHB  computer codes (without 
cranking and signature basis) designed for triaxial, axially symmetric and spherical shapes 
also exist \cite{Niksic2014_CPC185-1808}.

It turns out that without any assumptions on the form of the single-particle potential
self-consistent calculations generate single-particle properties (energies, alignments) 
which in general are similar to those obtained in phenomenological potentials. This is
illustrated in Fig.\ \ref{SHE-s-p} below which compares the single-particle spectra  of the 
superheavy $^{292}120_{172}$ nucleus obtained with phenomenological folded Yukawa (FY) 
potential,  non-relativistic Skyrme functionals SkP, SkM*, SLy6, SLy7, SkI1, SkI4 and Sk3, and 
CEDFs NL3, NL-Z, NL-Z2 and NL-VT1.  The Nilsson diagrams obtained in the RHB 
calculations with the CEDF NL3* (see Fig.\ \ref{s-p-def-254No}) show a lot of similarities with those 
obtained in the Woods-Saxon potential (see, for example,  Figs. 3 and 4 in Ref.\ \cite{CAFE.77}).
The behavior of the single-particle states in the rotating potential obtained in the
Woods-Saxon potential (see Fig. \ref{s-p-rot}) is very similar to that obtained in the CRMF
calculations (see Fig. 4 in Ref.\ \cite{ALR.98}). However, there is a principal difference in 
the physical  mechanisms related to the single-particle degrees of freedom between two types 
of  approaches and it is related to  time-odd mean fields \cite{DD.95,AR.00-to}. They are absent 
in phenomenological potentials but are present in self-consistent models. For example, in the 
CRHB framework they are related to the terms which break time-reversal symmetry, namely,  a 
magnetic potential ${\bm V}({\bm r})$ (see Eq.\ (\ref{Vmag})) and the the Coriolis term $-\Omega_x J_x$. 
In the DFT approaches, time-odd mean fields modify single-particle energies \cite{AA.10,SDMMNSS.10}, 
single-particle alignments  \cite{AR.00-to} and have large impact on rotational properties
of nuclei \cite{DD.95,AR.00-to}.

%%%%%%%%%%%%%%%%%%%%%%%%%%%%%%%%%%%%%%%%%%%%%
\section{Manifestation of independent particle motion in non-rotating and rotating nuclei}
%%%%%%%%%%%%%%%%%%%%%%%%%%%%%%%%%%%%%%%%%%%%%

    When considering the manifestations of independent particle motion one should 
separate the effects which emerge from the shell structure (as a coherent effect of motion of 
many particles) and those from individual motion of the  particles. The configuration-mixing 
interactions [which are accumulated in residual interaction term of the Hamiltitonian given by Eq.\ 
(\ref{Hamilt-mod})] such as pairing and the coupling to the low-lying collective vibrational degrees 
of freedom act destructively on the individual properties of the single-particle orbitals. In contrast,
the global effects emerging from the underlying shell structure are not that much affected by these 
residual interactions.   Thus,  a global shell structure at spin zero,  the superdeformation at high 
spin and the existence of superheavy nuclei are considered  as the examples of  such effects.

   Then the physical properties which sensitively depend on the single-particle features of 
individual orbitals are discussed. First, the energies of experimental and calculated 
one-quasiparticle states in non-rotating deformed nuclei are compared. Then, the analysis 
is extended to rotating nuclei at low and medium spins. The rotation acts as a tool to significantly 
reduce the role of pairing interaction \cite{SGBGV.89,NWJ.89,AF.05}.  Thus, nuclear systems 
at very high spins in which the impact of pairing is negligible are considered and the single-particle 
and  polarization effects due  to the occupation of  specific orbitals are analyzed. In addition, 
the phenomenon of band termination is discussed as an example of the competition of the
collective and single-particle degrees of freedom.

%%%%%%%%%%%%%%%%%%%%%%%%%
\subsection{Global shell structure at spin zero}
%%%%%%%%%%%%%%%%%%%%%%%%%

%%%%%%%%%%%%%%%%%%%%%%%%%%%%%%%%%%%%%%%%% 
\begin{figure}[t!]
\centering
\includegraphics[width=\linewidth]{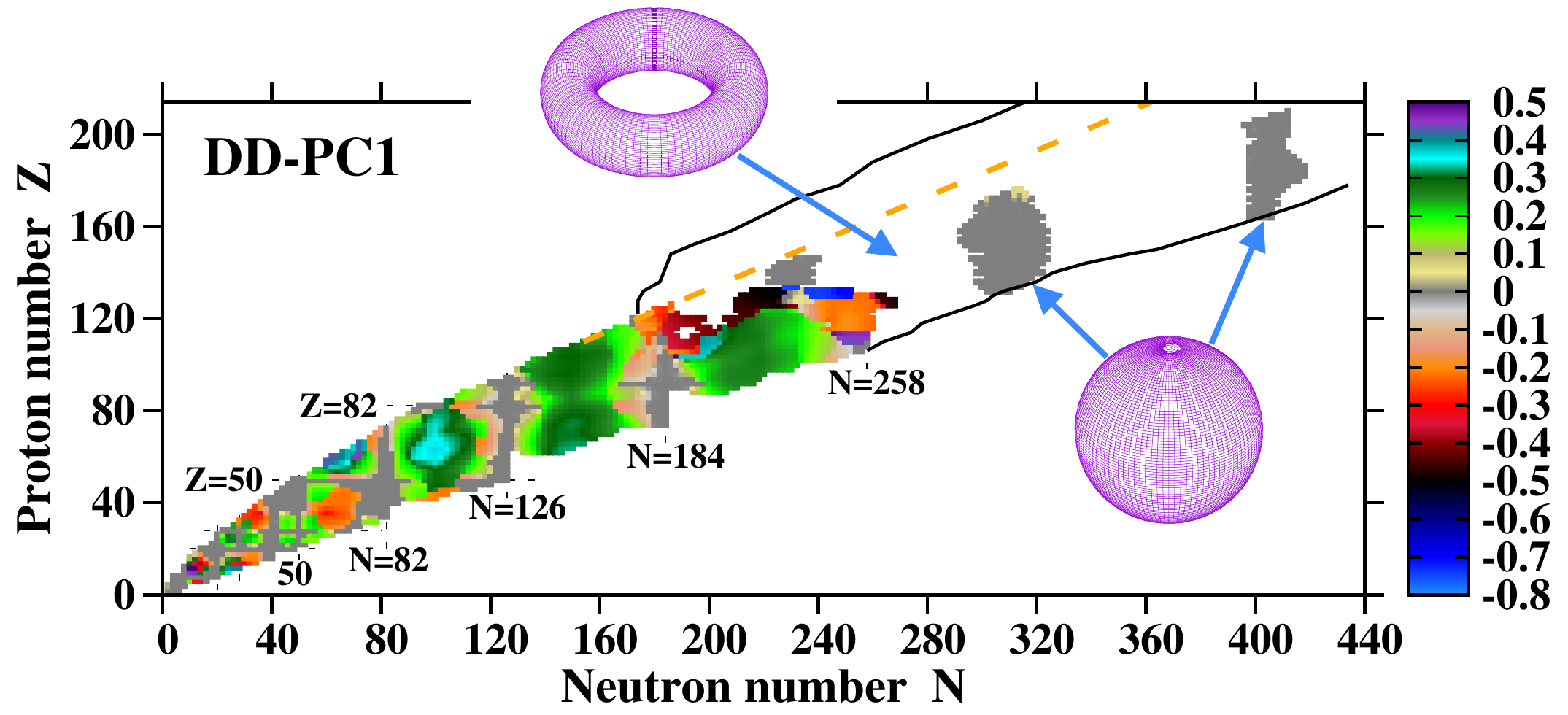}
\caption{Nuclear landscape at spin zero as obtained in the RHB calculations with the 
DD-PC1 functional. For proton numbers $Z\leq 130$, the ground states have ellipsoidal 
shapes. These nuclei are shown by the squares  the color of which indicates the 
equilibrium quadrupole deformation  $\beta_2$ (see colormap). Toroidal shapes dominate 
nuclear landscape for higher proton numbers (white region located between  two-proton 
and two-neutron drip  lines for toroidal nuclei shown by solid black lines). Figure taken 
from Ref.\ \cite{AA.21}.
\label{landscape}
}
\end{figure}
%%%%%%%%%%%%%%%%%%%%%%%%%%%%%%%%%%%%%%%%% 

  The state-of-the-art view on the nuclear landscape is shown in Fig.\ 
\ref{landscape}.  It is based on the results of the calculations presented in Refs.\ 
\cite{AA.21,AARR.14,AATG.19}. The bands (shown by gray color) of spherical nuclei 
along proton and neutron numbers 8, 20, 28, 50 and 82 (as well as for neutron 
number $N=126$) are seen in this figure.  They are due to large spherical shell 
gaps at these particle numbers which provide an extra stability of the nuclei. Note that 
these bands are more pronounced in neutron subsystem. Outside of these 
bands of spherical nuclei, the ground states of the $Z\leq 120$ nuclei are either oblate 
or prolate being in typical range of quadrupole deformations $|\beta_2| < 0.4$. Similar 
structures  and features are also obtained in non-relativistic calculations of Refs.\ 
\cite{MSIS.16,DGLGHPPB.10,Eet.12} but they  are  restricted to the nuclei below 
$Z \approx 120$. In general, existing experimental data confirms these model 
predictions \cite{SP.08} but there may be some differences between specific
model and experiment.

  With increasing proton number  these classical features disappear 
and only toroidal shapes are calculated as the lowest in energy.
This region (shown in white color between two black lines in Fig.\ \ref{landscape}) 
is penetrated only by  three islands (shown in gray color) of potentially stable 
spherical hyperheavy nuclei.  Their existence is due to substantial proton $Z = 154$ 
and 186 and neutron $N = 228$, 308, and 406 spherical shell gaps \cite{AA.21}
and substantial fission barriers around spherical minima \cite{AA.21,AATG.19}. 
However, these states are highly excited with respect of minima corresponding
to toroidal shapes. They could become the ground states if  relevant toroidal minima 
are unstable  with respect of so-called sausage deformations \cite{AA.21,Wong.73}.

    Thus, the richness of nuclear structure seen in experimentally known 
part  of  nuclear landscape  is replaced by a more uniform structure of the nuclear 
landscape in the region  of hyperheavy ($Z\geq 126)$ nuclei dominated by toroidal 
nuclei  \cite{AA.21,AATG.19}. This transition from compact ellipsoidal-like shapes 
to non-compact toroidal shapes is driven by the enhancement of the role of Coulomb 
interaction with increasing $Z$ \cite{AATG.19}. Thus, the former shapes become 
either  unstable against fission or energetically unfavored in hyperheavy nuclei.

    Single-particle degrees of freedom play also an important role in the definition of 
the boundaries of nuclear landscape.  For example, the position of two-neutron drip line 
of ellipsoidal nuclei depends sensitively on the single-particle energies of high-$j$ states 
located in the vicinity of neutron continuum threshold \cite{AARR.15}. The transition from 
ellipsoidal to toroidal shapes  drastically modifies the underlying single-particle structure and 
as a result lowers  the energy of the Fermi level for protons \cite{AA.21}. As a consequence, 
a substantial shift of  the two-proton drip line from its expected position for ellipsoildal shapes 
(shown by  dashed orange line in Fig.\ \ref{landscape}) towards more proton-rich nuclei for 
toroidal shapes (shown by solid  line in Fig.\ \ref{landscape}) takes place \cite{AA.21}.

%%%%%%%%%%%%%%%%%%%%%%%%%
\subsection{Superheavy nuclei}
%%%%%%%%%%%%%%%%%%%%%%%%%

    With increasing proton number beyond $Z=100$ the fission barriers provided
by the liquid drop become rather small and then for higher $Z$ they disappear (see Fig.\ 
10.6 in Ref.\ \cite{NilRag-book}).  It turns out that the existence of the heaviest nuclei 
with $Z > 104$ is primarily determined by shell effects due to the quantum-mechanical
motion of protons and neutrons inside the nucleus 
\cite{SGK.66,Meld-67,OU.15,GMNORSSSS.19,AALMZ.21}. In these nuclei the
heights of fission barriers are entirely determined by the shell corrections and 
they would not exist without shell effects. These effects  also play a central role for 
the production, stability and spectroscopy of superheavy nuclei.

%%%%%%%%%%%%%%%%%%%%%%%%%%%%%%%%%%%%%%%%% 
\begin{figure}[t!]
\centering
\includegraphics[width=5.8cm,]{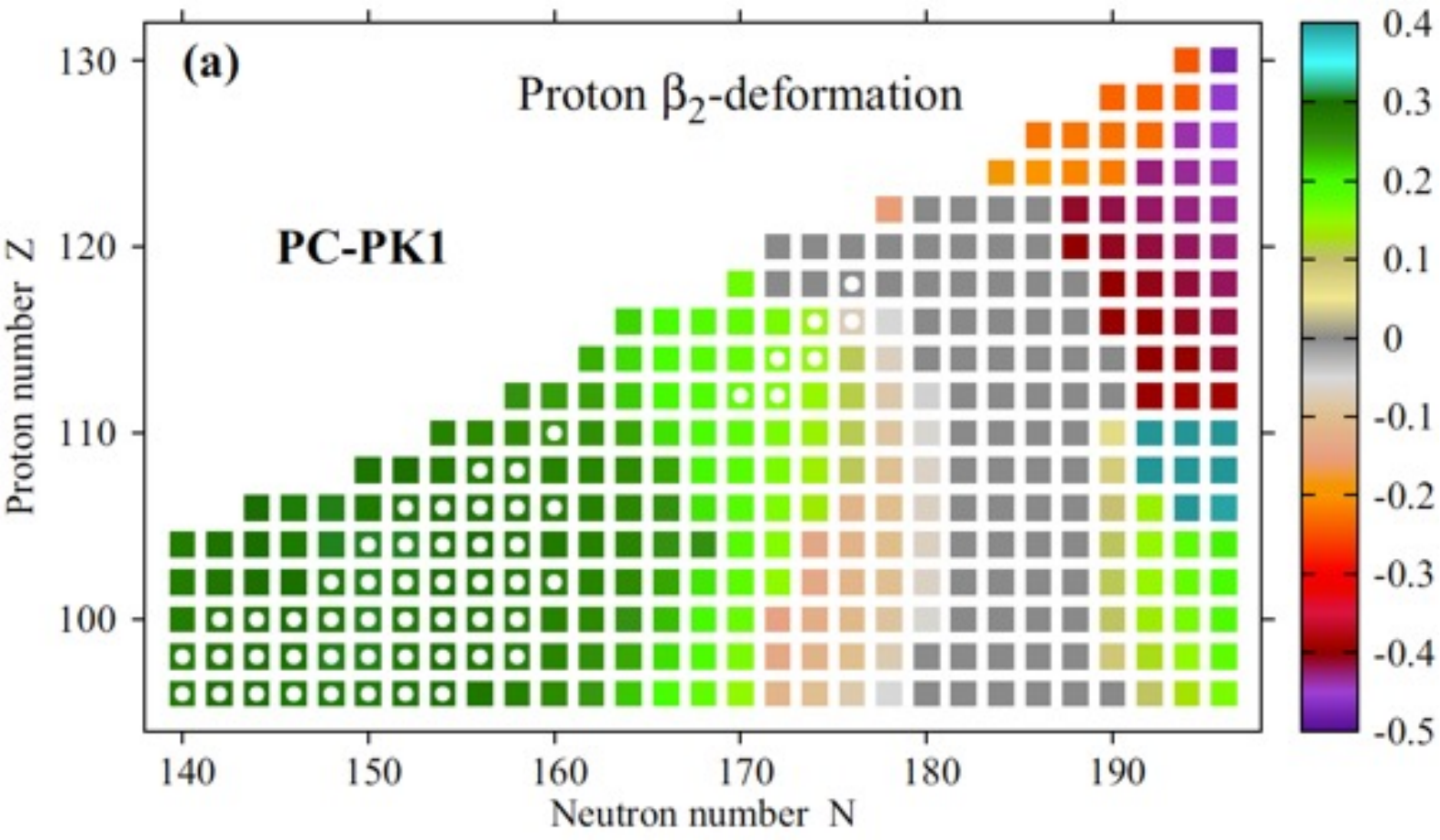}
\includegraphics[width=5.8cm]{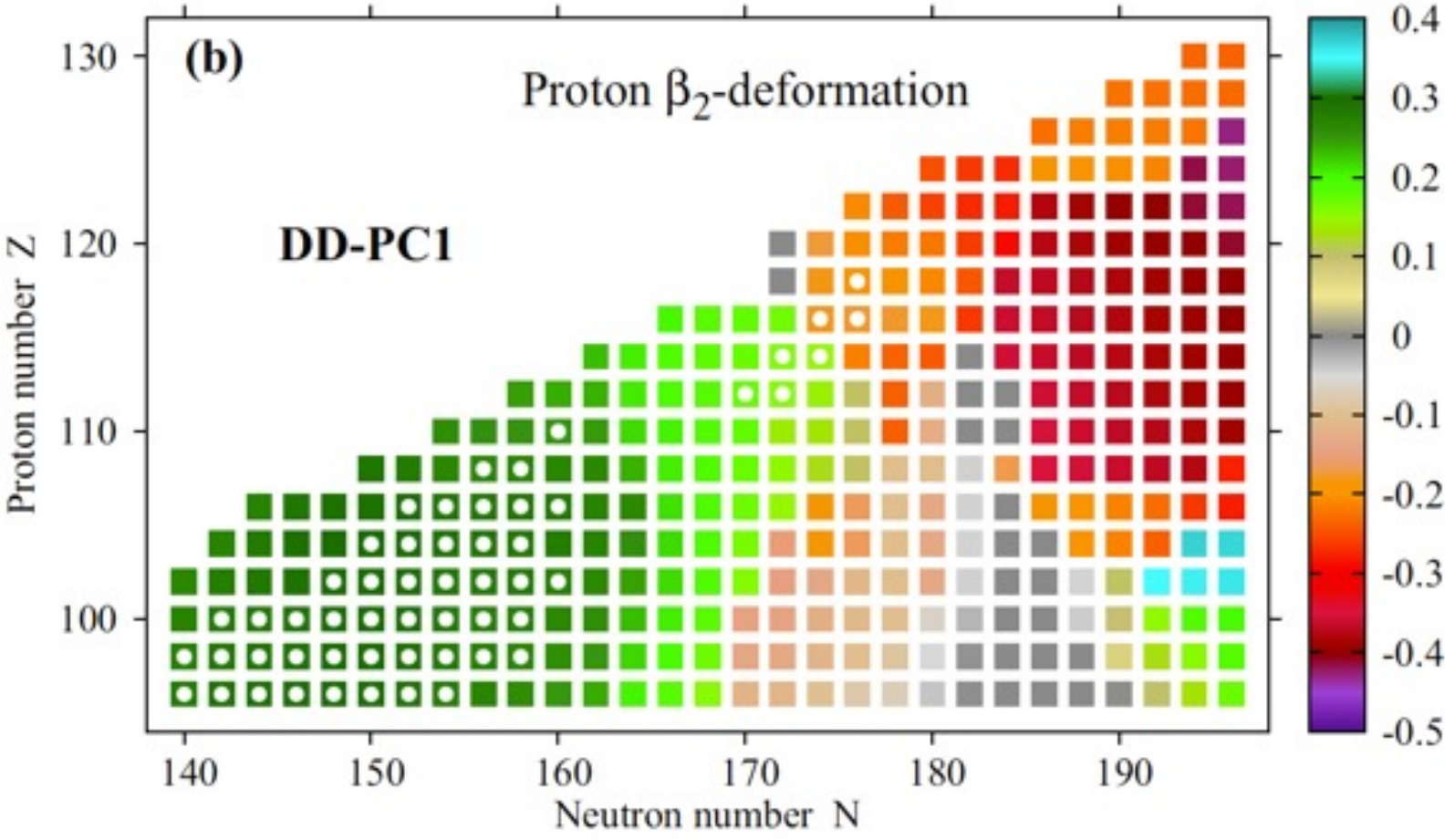}
\caption{Charge quadrupole deformations $\beta_2$ of even-even superheavy
nuclei obtained in the RHB calculations with indicated functionals. Experimentally known 
nuclei are indicated by white circles. Figure taken from Ref.\  \protect\cite{AANR.15}. 
\label{deformations-SHE}
}
\end{figure}
%%%%%%%%%%%%%%%%%%%%%%%%%%%%%%%%%%%%%%%%% 

   Although state-of-the-art theoretical models provide a reasonable description of 
many aspects of the physics of superheavy nuclei, they face substantial challenges 
in the prediction of the location of next spherical shell closures 
\cite{BRRMG.99,SP.07,OU.15,GMNORSSSS.19,AALMZ.21,AANR.15}.  These challenges are 
illustrated in Fig.\ \ref{deformations-SHE} which shows the map of calculated ground state 
quadrupole deformations obtained with two covariant energy density functionals. The predictions 
of these two functionals are drastically different in the vicinity of the $Z=120$ and 
$N=184$ lines. The PC-PK1 functional predicts wide  bands of spherical nuclei in the $(Z,N)$ 
chart along $Z=120$ and $N=184$. In contrast, in the calculations with DD-PC1, the band 
along $Z=120$ does not exist and narrower band along $N=184$ is seen only for 
$Z\leq 114$. These discrepancies are due to modest differences in the single-particle properties 
of these functionals and related differences in the competition of shell effects at 
spherical and deformed shapes  \cite{AANR.15}. Note that the width of the band of 
spherical nuclei in the $(Z,N)$ chart along a specific particle number corresponding 
to a shell closure indicates the impact of this shell closure on the structure of neighboring 
nuclei.
 
%%%%%%%%%%%%%%%%%%%%%%%%%%%%%%%%%%%%%%%%% 
\begin{figure}[t!]
\centering
\includegraphics[width=8.0cm]{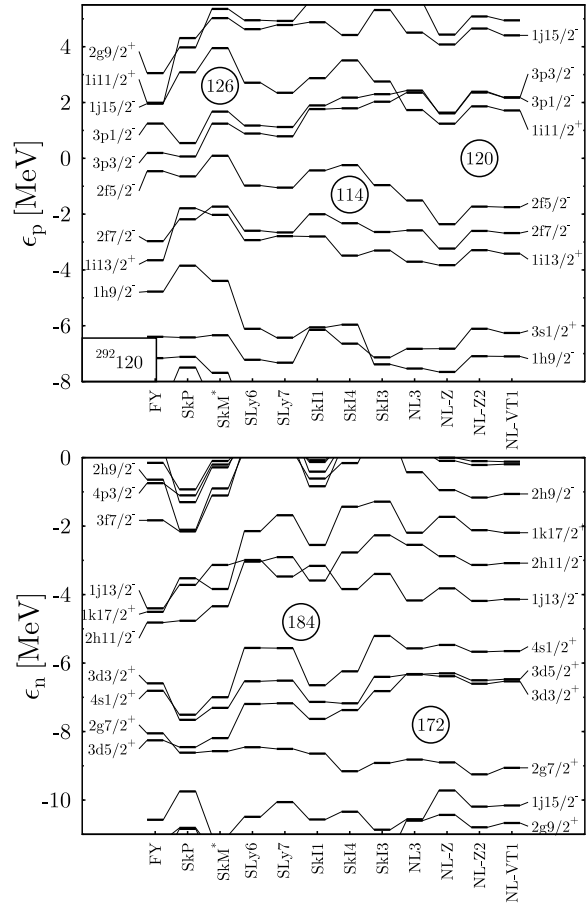}
\caption{Single-particle energies of proton (top) and neutron (bottom)
states in the $^{292}120_{172}$ nucleus obtained at spherical shape
with indicated non-relativistic and relativistic functionals. Figure taken 
from Ref.\  \protect\cite{BRRMG.99}. 
\label{SHE-s-p}
}
\end{figure}
%%%%%%%%%%%%%%%%%%%%%%%%%%%%%%%%%%%%%%%%% 
 
   There is  a wide variety of the predictions for proton and neutron spherical 
shell closures in superheavy nuclei obtained in different models. 
These are proton numbers at $Z=114, 120$ and 126 and neutron numbers at $N=172$ 
and 184 (see Fig.\ \ref{SHE-s-p} and Refs.\ \cite{BRRMG.99,SP.07,GMNORSSSS.19,AANR.15}. 
However,  in the CDFT the $N=172$ could be eliminated as a potential candidate when the 
deformation effects are taken into account \cite{AANR.15} and the $Z=120$ nuclei become 
deformed in a number of functionals when both deformation and correlations beyond mean 
field effects  are considered \cite{SALM.19}.  The challenges the models face in the prediction 
of spherical shell closures are due to two factors. At present, self-consistent models do not 
provide a spectroscopic quality of the description of the single-particle spectra \cite{DABRS.15}. 
In  contrast, the phenomenological potentials better describe single-particle spectra in known
nuclei (see Ref.\ \cite{SP.07} and references therein) but they fail to incorporate self-consistency 
effects (such  as a depletion of the density in central region of nucleus) which are important in 
superheavy nuclei (see discussion of Fig.\ \ref{density-bubble}).

    Fig.\ \ref{deformations-SHE} illustrates that only for a few experimentally
known $Z=116$ and 118 nuclei the CDFT predictions differ. In general, with possible
exception of these  high-$Z$ isotopic chains, existing experimental data on superheavy 
nuclei are described with comparable level of accuracy by all existing models 
\cite{GMNORSSSS.19,AALMZ.21,AANR.15}. This is undeniable success of independent  
particle model the global consequences of which (shell structure) led to the predictions of 
superheavy nuclei in 1960s within relatively simple models \cite{SGK.66,Meld-67}. Experimental 
data collected over these years fully confirmed these predictions and some differences in the
predictions of next spherical shell closures are secondary to this success. This field
or research (both experimental and theoretical) remains extremely active which is 
illustrated by recent reviews \cite{OU.15,GMNORSSSS.19,AALMZ.21}.  However, at present 
available  experimental data does not allow to give a preference to the predictions of one or another 
model and to decide where next (if any) proton and neutron spherical shell  closures are located  
or whether there is an island of spherical superheavy nuclei.

%%%%%%%%%%%%%%%%%%%%%%%%%
\subsection{Superdeformation at high spin}
%%%%%%%%%%%%%%%%%%%%%%%%%

    Fig.\ \ref{landscape} shows that the ground states of the nuclei in experimentally known part 
of the nuclear chart are characterized by quadrupole deformation $\beta_2$  which is located 
typically in the range $-0.35 \leq \beta_2 \leq +0.35$.  Strutinsky has predicted excited minimum 
with $\beta_2 \sim 0.6$ in 
the second (superdeformed, SD) potential well of potential energy curves of deformed nuclei
in Ref.\ \cite{Strut67}. At low spin such minima are located at high excitation energies, but they can be 
brought down to the yrast line by fast rotation since it favors extremely deformed shapes.  This was 
confirmed later in the mic+mac  calculations of Ref.\ \cite{Ander.76} which predicted the doubly magic SD 
band in $^{152}$Dy with 2:1 semi-axis ratio.  The existence of the SD bands is due to the shell effects 
associated with  large proton and neutron SD shell gaps. For example, in the $A\sim 150$ region
of superdeformation these shell effects are produced by large proton $Z=66$ and neutron $N=86$  
SD shell  gaps which exist both in  phenomenological potentials (see Fig.\ \ref{s-p-rot} in this paper 
and Refs.\ \cite{NWJ.89,Rag.93}) and in the DFT calculations \cite{ALR.98}.

  The phenomenon of superdeformation at high spin has been confirmed experimentally by the 
observation of the first SD band in $^{152}$Dy in 1986 \cite{Twin}. From that time, a significant
amount of experimental data on the SD bands in different mass regions  has been collected
\cite{Firestone.2002}. The mic+mac and DFT-based cranked  approaches rather well describe 
experimental observables such as dynamic $J^{(2)}$ and kinematic $J^{(1)}$ moments of inertia and
transitional quadrupole moments $Q_t$ of these bands (see Refs.\ \cite{CRHB,ALR.98,NWJ.89,Rag.93}
and references quoted therein). The calculations also reveal a substantial dependence of 
these physical observables on the occupation of high-$N$ intruder orbitals; here $N$ stands 
for the principal quantum number of dominant component of the wave function. For example, they
strongly depend on the number of occupied $N=6$ protons and $N=7$ neutrons in the
$A\sim 150$ region of superdeformation. This region is also of special interest  since pairing
is negligible  in the majority of the SD bands. As a consequence, it provides one of the best 
examples of independent particle motion in nuclear physics (see detailed discussion in the 
last section).

%%%%%%%%%%%%%%%%%%%%%%%%%%%%%
\subsection{The phenomenon of band termination}
%%%%%%%%%%%%%%%%%%%%%%%%%%%%%

%%%%%%%%%%%%%%%%%%%%%%%%%%%%%%%%%%%%%%%%% 
\begin{figure}[t!]
\centering
\includegraphics[width=8.0cm]{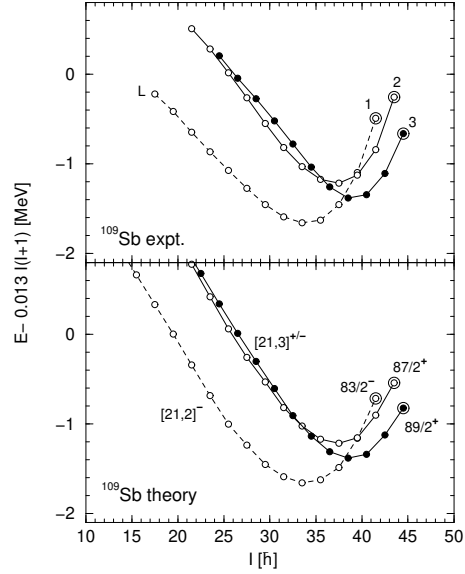}
\vspace{-1.0cm}
\caption{The comparison of experimental  and calculated $E-E_{RLD}$ curves 
for  the  bands $1-3$ in $^{109}$Sb. The energies have been normalized in such a way that
calculated and experimental curves for the band 1 have the same value at the minimum of 
the $E-E_{RLD}$ curve. The energies of the unlinked experimental bands 2 and 3 are chosen 
so that their $E-E_{RLD}$ minima have the same energies as theoretical counterparts. 
Terminating states are indicated by large open circles and their spins are displayed. 
Solid and dashed lines are used for 
the configurations with $\pi=+$ and $\pi=-$, respectively. The states with signature
$\alpha_{tot}=+1/2$ and $\alpha_{tot}=-1/2$ are shown by solid an open symbols, respectively.  Figure 
taken  from Ref.\ \cite{PhysRep-SBT}.
\label{E-ERLD}
}
\end{figure}
%%%%%%%%%%%%%%%%%%%%%%%%%%%%%%%%%%%%%%%%% 

  One of clear manifestations of the independent particle motion is the phenomenon 
of band termination \cite{PhysRep-SBT,HFMSAD.72,WKW.80}. It definitely reveals the 
fact that each single-particle orbital occupied in nucleonic configuration possesses a limited 
and state-dependent angular momentum.   Of special interest are so-called {\it smooth 
terminating bands} which 
show a continuous transition from high collectivity  at low and medium spin values to
a pure particle-hole (terminating) state at the maximum spin which can be built 
within the configuration \cite{PhysRep-SBT,RJFSW.95}. Note that this feature of finite multi-fermion 
systems has so far only been observed and studied in atomic nuclei. In the terminating state, the symmetry axis 
coincides with the rotation axis.  Since collective rotation along the symmetry axis is forbidden 
in quantum mechanics, no further  angular momentum can be brought into the system with 
the same occupation of single-particle  orbitals and thus this state represents the termination 
of the rotational band.

   The phenomenon of band termination has been observed in several regions of nuclear chart 
(see review in Ref.\ \cite{PhysRep-SBT}). However, the best examples of smooth terminating
bands are seen in the $A\approx 110$ region in which the nuclei have several valence particles
and holes outside the $^{100}$Sn core. 
%I consider here the classical example of terminating
%bands in $^{109}$Sb \cite{RJFSW.95,PhysRep-SBT}  in order to illustrate the major features of 
%smooth band termination. 
A classical example of terminating
bands in $^{109}$Sb \cite{PhysRep-SBT,RJFSW.95} is considered here
in order to illustrate the major features of 
smooth band termination. 
A critical feature of these bands is the fact that their dynamic moments 
of inertia $J^{(2)}$ gradually decrease with increasing rotational frequency to unusually low
values (near 1/3 of rigid-body value) at the highest observed frequencies. This is definite indication
of significant suppression of pairing correlations \cite{PhysRep-SBT}  making this type of rotational
structures as one of the best examples of independent particle motion. 

   Fig.\ \ref{E-ERLD}  compares the experimental and calculated energies $E$ of rotational 
bands with respect of rigid rotor reference $E_{RLD} = A I(I+1)$, where $A$ is the moment of inertia 
parameter. The calculations are performed in configuration-dependent cranked Nilsson-Strutinsky
approach (CNS) approach \cite{PhysRep-SBT,BR.85}. The configurations relatively to the 
$^{100}$Sn core are labelled using the shorthand  notation  $[p_{1}\,p_{2},n]^{\alpha_{tot}}\equiv 
[\pi (g_{9/2})^{-p_{1}}\,\,(h_{11/2})^{p_{2}}\,\,
(g_{7/2}\,\,d_{5/2})^{Z-50+p_{1}-p_{2}}\,\,\otimes
\nu (h_{11/2})^{n} (g_{7/2}\,\,d_{5/2})^{N-50-n}]^{\alpha_{tot}}$,
where $\alpha_{tot}$ is the total signature of the configuration (only sign is shown)
and $p_{1}$, $p_{2}$ and $n$ are the numbers of proton holes in
the $g_{9/2}$ orbitals, of protons in the $h_{11/2}$ orbitals and
of neutrons in the $h_{11/2}$ orbitals, respectively.  One can see that the CNS calculations without
pairing very well reproduce experimental data. Some discrepancies seen at low $I\leq 20\hbar$ spin 
are due to the neglect of pairing.

\begin{figure}[t!]
\centering
\includegraphics[width=6.8cm]{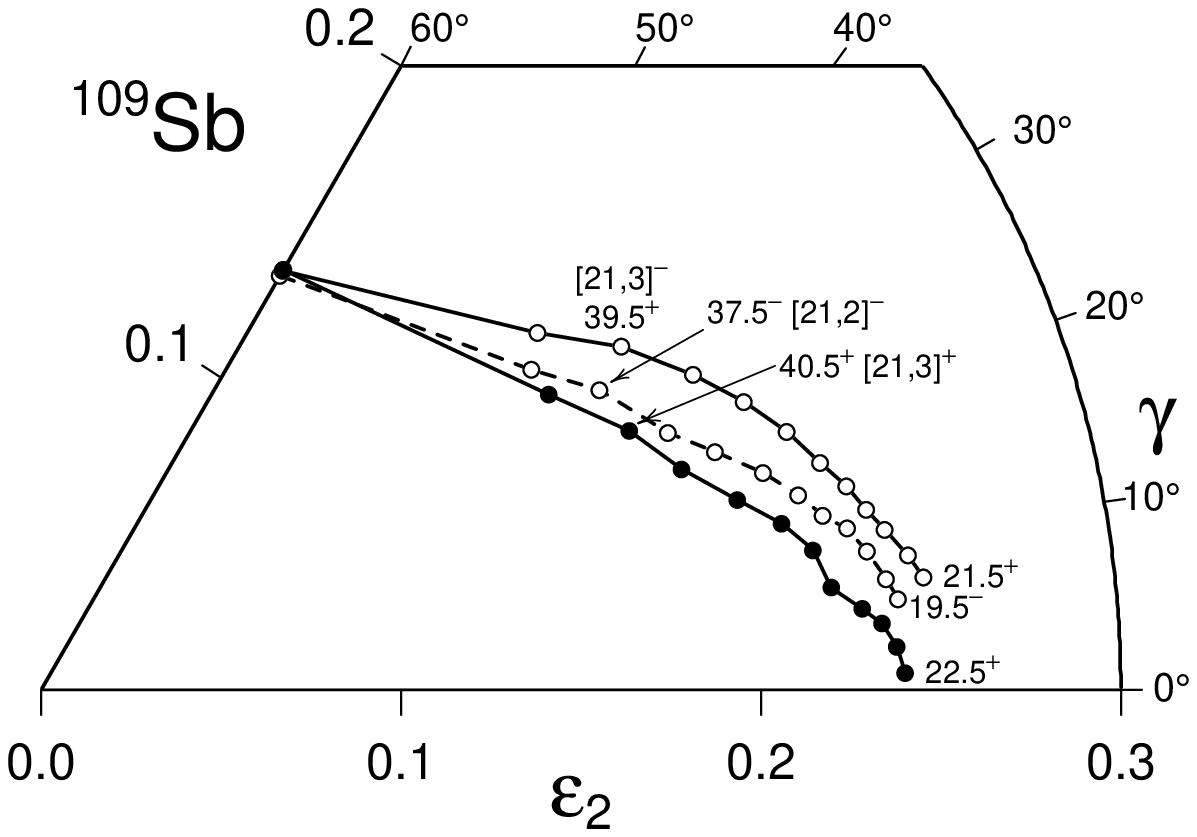}
%\hspace{-1.0}
\includegraphics[width=4.7cm]{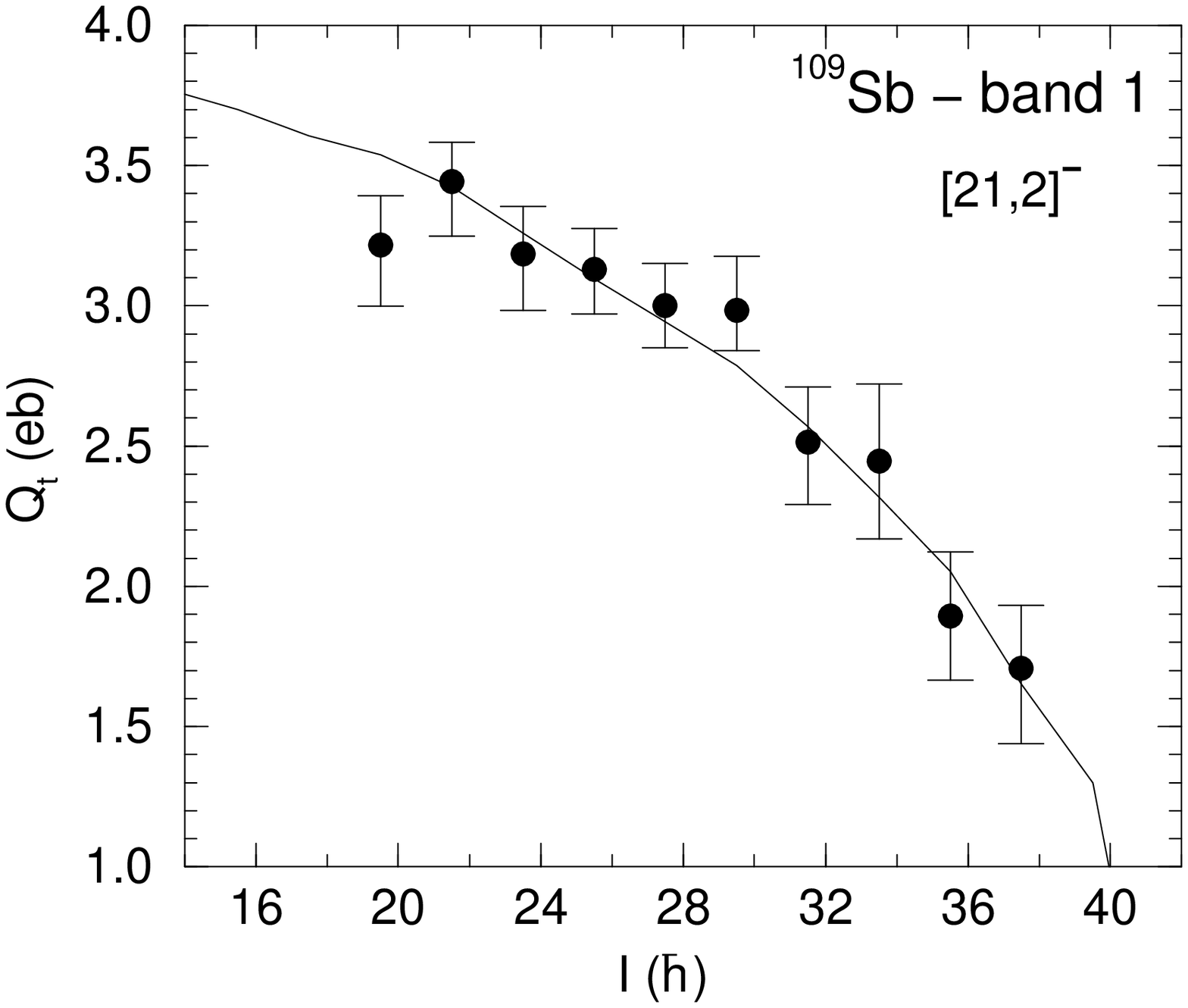}
\caption{(left panel) Calculated deformation paths in the $(\varepsilon_2, \gamma$)
plane of the configurations assigned to smooth terminating bands 1-3 in $^{109}$Sb.
The spins of some states are shown.
(right panel) The comparison of experimental and calculated transition quadrupole 
moments $Q_t$ of the band 1 and assigned theoretical configuration. Figure taken 
from Ref.\  \cite{PhysRep-SBT}.
\label{109Sb-deform}
}
\end{figure}

   The CNS calculations suggest that rotational bands of interest are near prolate
at low spin and thus they involve collective rotation about an axis perpendicular 
to the symmetry axis. With increasing spin the valence nucleons gradually align 
their spin vectors with the axis of rotation via the Coriolis interaction. This  causes 
the nuclear shape to gradually trace a path through the triaxial ($\gamma$) plane, 
toward the non-collective oblate shape at $\gamma=+60^{\circ}$ (see left panel
of Fig.\ \ref{109Sb-deform}).  After the available spin is exhausted, consistent with 
the Pauli principle, the band terminates.  This gradual change from collective near-prolate  
$(\gamma\sim 0^{\circ})$ to non-collective oblate ($\gamma=+60^{\circ}$) shape leads to 
a gradual decrease of transition quadrupole moment $Q_{t}$ which agrees with available 
experimental data (see right panel in Fig.\ \ref{109Sb-deform}). The termination spins, which 
depend on the configuration, are well defined property for terminating bands and they 
confirm the termination process. For example, the detailed structure of the $[21,2]^-$ (band 1)
terminating $I=\frac{83}{2}^-$ state in $^{109}$Sb is $\pi (g_{9/2})^{-2}_{8} (g_{7/2} d_{5/2})^{2}_{6} (h_{11/2})^{1}_{5.5}
\otimes \nu (g_{7/2} d_{5/2})^{6}_{12} (h_{11/2})^{2}_{10} $. Note that fully self-consistent
cranked relativistic mean field calculations confirm this physical picture \cite{VALR.05}. However, 
due to technical reasons  it is difficult to follow smooth terminating bands  up to their terminating 
states in  such calculations.

%%%%%%%%%%%%%%%%%%%%%%%%%%%%%%%%%%%%%%%%% 
%%%%%%%%%%%%%%%%%%%%%%%%%%%%%%%%%%%%%%%%% 

%%%%%%%%%%%%%%%%%%%%%%%%%%%%%%%%%%%
\subsection{Single-particle states in deformed nuclei}
%%%%%%%%%%%%%%%%%%%%%%%%%%%%%%%%%%%

%%%%%%%%%%%%%%%%%%%%%%%%%%%%%%%%%%%
\subsubsection{Non-rotating nuclei}
%%%%%%%%%%%%%%%%%%%%%%%%%%%%%%%%%%%

   The detailed comparison of experimental and theoretical  information on 
the properties of specific individual states can shed additional light on the 
validity in independent particle motion in atomic nuclei. In non-rotating nuclei, 
such information is provided by the energies of the single-particle states,  
their densities and by the transitions between the single-particle states.
However, such densities are not accessible experimentally and calculated 
transition probabilities are prone to substantial theoretical errors.  One 
could imagine that spherical nuclei would provide the cleanest and simplest
set of data on the single-particle properties.  It turns out that this is not a case
because of substantial residual interaction due to (quasi)particle-vibration 
(quasiparticle-phonon) coupling.  As a consequence, the wavefunctions
of the states in the spherical nuclei are not of pure single-particle nature
since they are substantially polluted by the vibrational admixtures
\cite{MBBD.85,AL.15,CCSB.14}.

%%%%%%%%%%%%%%%%%%%%%%%%%%%%%%%%%%%%%%%%% 
\begin{figure}[t!]
\centering
\includegraphics[width=5.58cm]{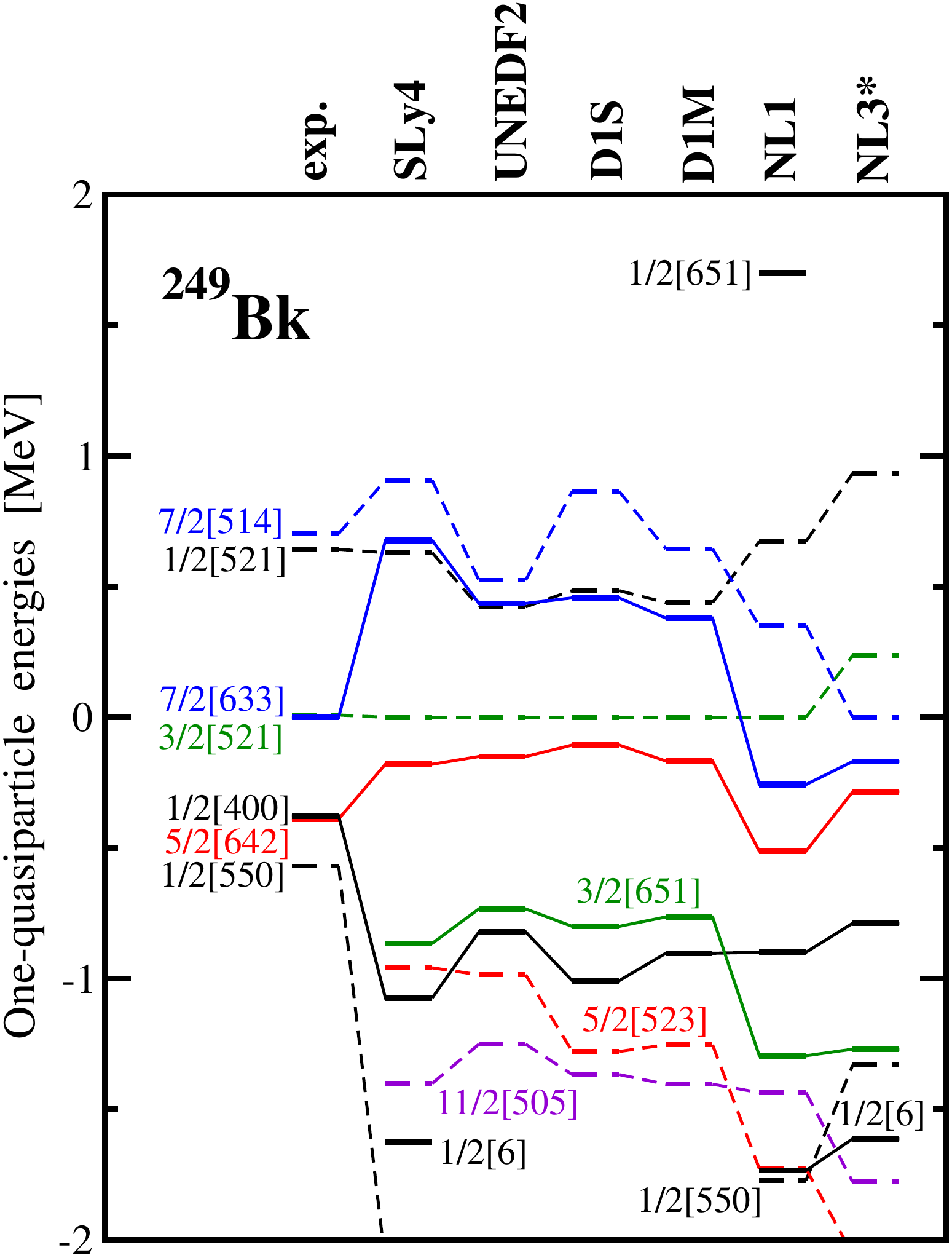}
\includegraphics[width=5.0cm]{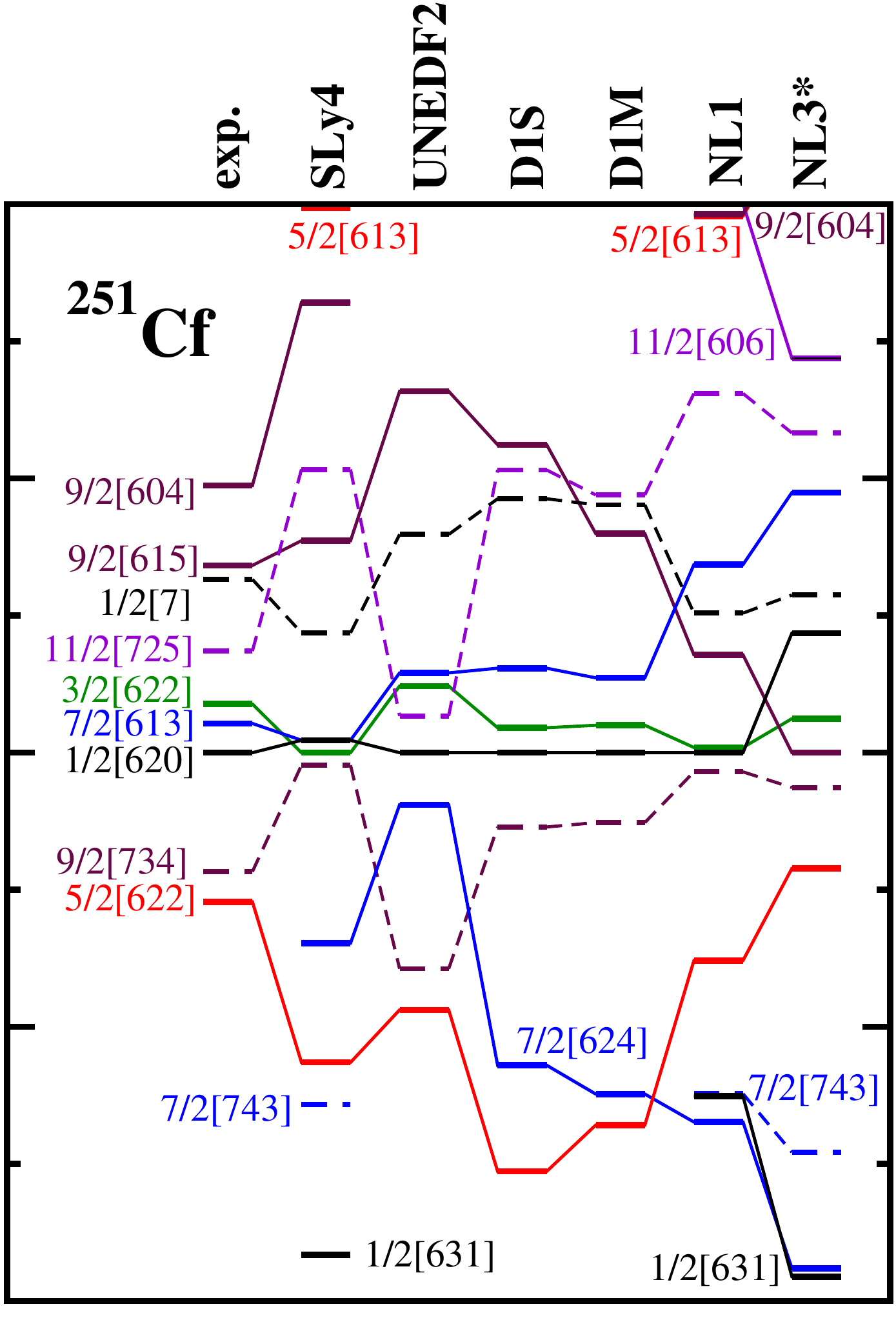}
\caption{Experimental and calculated quasiparticle spectra in $^{249}$Bk 
and $^{251}$Cf.  Solid and dashed  lines are used for  positive- and negative-parity 
states, respectively. The states are labeled by the Nilsson label of the dominant 
component of the wave function when the squared amplitude of this component exceeds 
50\%. Otherwise, they are labeled by $\Omega[N]$ where $N$ is principal quantum 
number of the components of wave function whose cumulative contribution into the wave 
function is dominant.  Figure taken from Ref.\ \cite{DABRS.15}.
\label{s-p-def}
}
\end{figure}
%%%%%%%%%%%%%%%%%%%%%%%%%%%%%%%%%%%%%%%%% 

   The impact of the quasiparticle-vibration coupling is reduced in 
deformed nuclei since the part of  such correlations is accounted at the level of
deformed mean field. Indeed, the admixtures of the phonons to the
structure of the ground and low-lying states in deformed rare-earth
and actinide odd-mass nuclei is relatively small (especially, when 
compared with spherical open shell nuclei \cite{AL.15}) according to the 
quasiparticle-phonon model \cite{ABNSW.88,SSMJ.15}. Thus, deformed
nuclei can provide a better and cleaner examples of independent particle
motion.

%%%%%%%%%%%%%%%%%%%%%%%%%%%%%%%%%%%%%%%%% 
\begin{figure}[t!]
\vspace{-1.5cm}
\centering
\includegraphics[width=5.58cm]{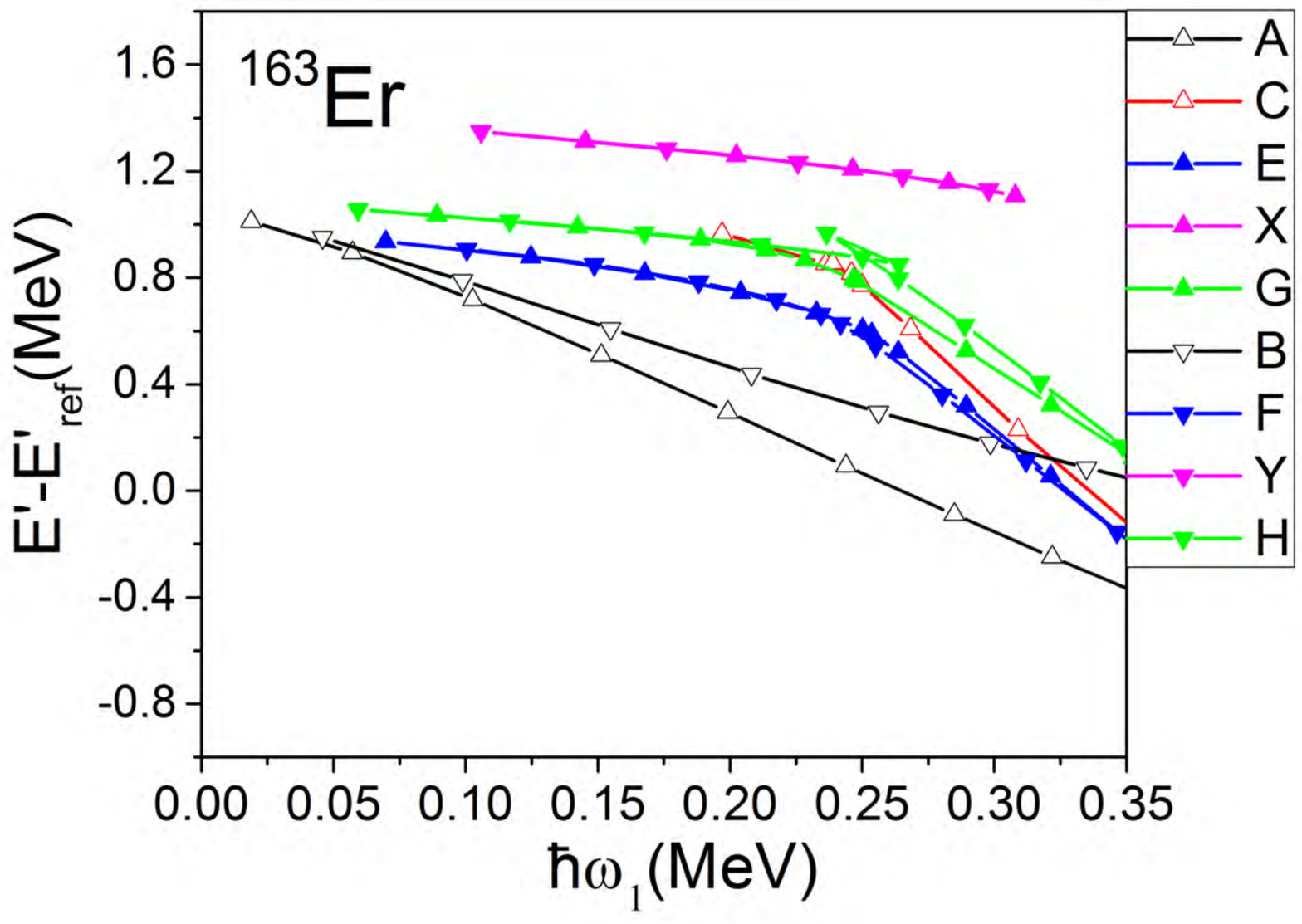}
\includegraphics[width=5.58cm]{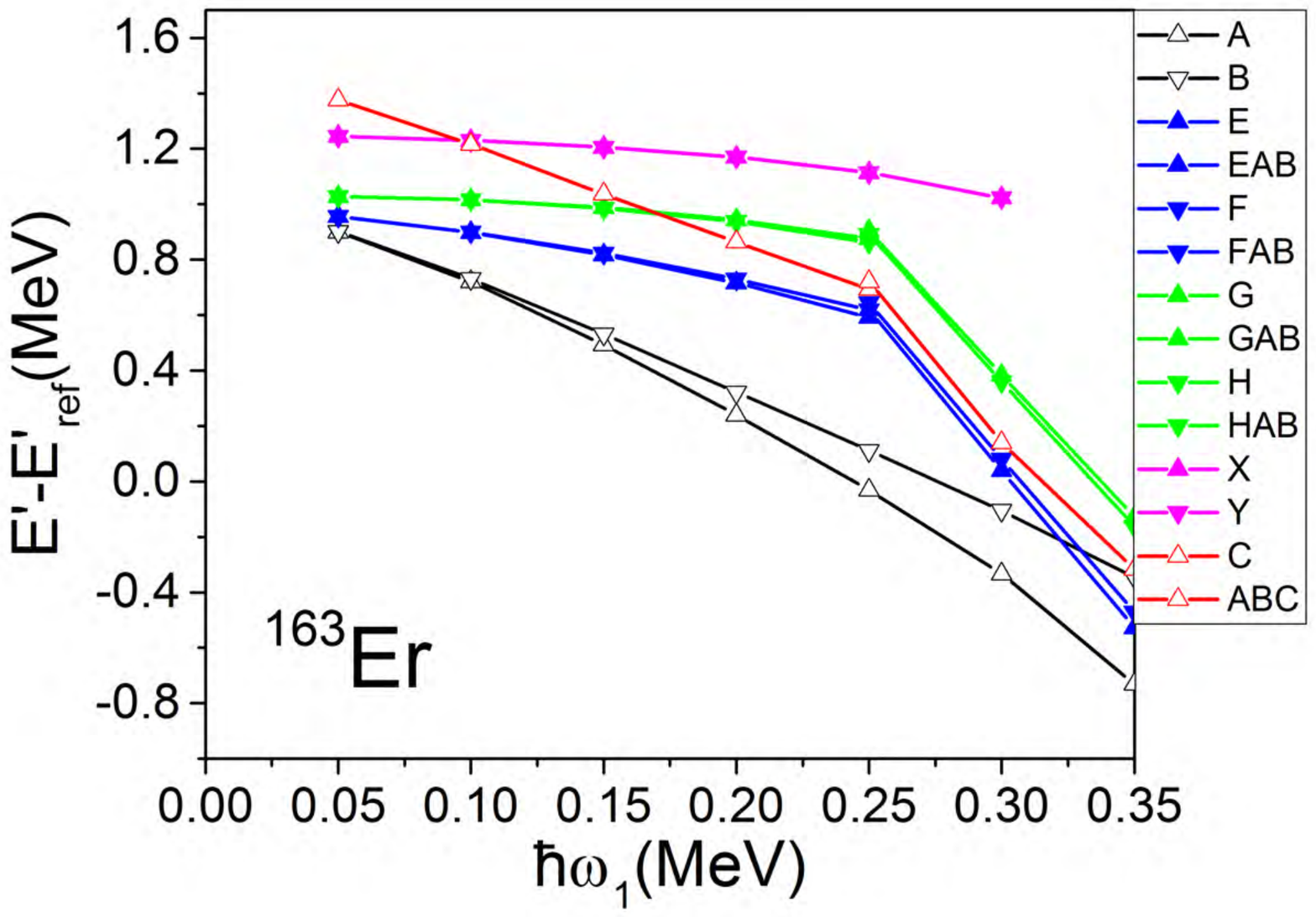}
\vspace{-1.5cm} 
\caption{Experimental (left panel) and calculated (right panel)  
Routhians relative to the $g$-band reference Routhian $E_{ref}$ as a function
of rotational frequency $\hbar \omega_1$.  Note that 
the data are restricted to the simplest configurations since two-quasiproton and 
several three-quasineutron configurations are left away.
The convention for $(\pi, \alpha)$ is the following: open and solid symbols
are used for $\pi=+$ and $\pi=-$, respectively, and the $\alpha=+1/2$ and
$\alpha=-1/2$ states are shown by triangles up and triangles down, respectively.
Figures taken from Ref.\ \cite{F.18}.
\label{s-p-routh}
}
\end{figure}
%%%%%%%%%%%%%%%%%%%%%%%%%%%%%%%%%%%%%%%%% 

    Fig.\ \ref{s-p-def}  compares experimental and calculated energies 
of one-quasiparticle states $E^{qp}_i$ in odd-$A$ $^{249}$Bk and $^{251}$Cf
nuclei. The results of the calculations with non-relativistic Skyrme (SLy4, UNEDF2), 
Gogny (D1S, D1M) and covariant (NL1, NL3*) energy density functionals are 
shown in this figure. These are fully self-consistent calculations which means that the total energies 
$E_i$ of the nucleonic  configurations with blocked  $i$-th  single-particle state of 
interest are obtained within Hartree-Fock-Bogoliubov or RHB frameworks 
\cite{RS.80}.  Then the ground state is associated with nucleonic configuration 
which has the lowest energy $E_{lowest}$ and the energies of the excited states $E^{qp}_i$ 
are defined as $E^{qp}_i = E_i - E_{lowest}$. Note that in Fig.\ \ref{s-p-def}  the 
particle  and hole states are plotted at $E^{qp}_i$ and 
$-E^{qp}_i$, respectively. This is done in order to facilitate the comparison of these
self-consistent results with the Nilsson diagrams.

  One can see that the energies of some deformed one-quasiparticle states are 
rather  well described  in specific functionals, but others deviate appreciably from the 
experiment.  The detailed comparison of experimental data with calculations is presented in 
Ref.\ \cite{DABRS.15}. Considering an average level of the accuracy of the description of 
experimental data, it is difficult to give a clear preference to one or another functional.  At 
present,  a systematic analysis of the accuracy of the reproduction of the single-particle 
spectra in deformed  nuclei is available only in the RHB framework \cite{AS.11} and 
smaller in scope (only for the Rb, Yb  and Nb isotopic chains) analysis is carried out 
in the HFB calculations with Gogny forces  (see Ref.\ \cite{RSR.11} and references therein).  
These investigations reveal two sources of inaccuracies in the description of the energies of 
the single-particle states, namely, low effective mass leading to a stretching of the energy 
scale of the calculated results as compared with experimental ones and incorrect relative 
positions of some single-particle states \cite{DABRS.15,AS.11}. On the absolute scale these 
deficiencies are not large especially considering the fact that no (or very limited) information 
on single-particle degrees of freedom has been taken into account in the fitting protocols of 
covariant (non-relativistic) energy density functionals.  The accounting  of quasiparticle-vibration 
coupling (QVC) improves the agreement with experiment by both improving the description of 
the energies of individual states and increasing the density of the single-particle states in the 
vicinity of the Fermi level. This was illustrated for the $^{251}$Cf nucleus in the RHB+QVC 
framework in  Ref.\  \cite{ZBNLRS.22}.

   In general, phenomenological potentials provide better description of the energies of the 
single-particle states  in deformed odd-$A$ nuclei because they are fitted to this type of experimental  
data \cite{CDNSW.87,BDNO.89,BR.85}. However, such accuracy is obtained at the cost of 
neglect of self-consistency  effects \cite{AS.11} and effective incorporation of the effects of 
particle-vibration coupling \cite{ZBNLRS.22}. 

%%%%%%%%%%%%%%%%%%%%%%%%%%%%%%%%%%%
\subsubsection{Rotating nuclei in the pairing regime}
%%%%%%%%%%%%%%%%%%%%%%%%%%%%%%%%%%%

   As discussed earlier the rotation of the nuclei can generate substantial modifications 
of the  single-particle energies and provides single-particle alignments $\left< j_x \right>_i$ as
a new measure of the single-particle properties.  This leads to a new and very robust probes of 
the single-particle structure since some pairs of the orbitals emerging from a given non-rotating 
state show significant signature splitting (see, for example, the $[770]1/2(\alpha=\pm 1/2)$ and 
$[761]3/2(\alpha=\pm1/2)$ pairs of the orbitals in Fig.\ \ref{s-p-rot}) while other pairs (such 
as $[532]5/2(\alpha=\pm1/2)$ and $[633]7/2(\alpha=\pm1/2)$ in Fig.\ \ref{s-p-rot}) show
no signature splitting. 

    These features can be used for a reliable interpretation of experimental data. This 
is demonstrated  in Fig.\ \ref{s-p-routh} which compares  experimental Routhians with calculated 
ones for selected set of rotational bands in odd-neutron  $^{163}$Er nucleus. The excitation 
energies are taken relative to a reference configuration which is the ground state ($g$-) configuration 
in even-even nucleus. Experimental Routhian is well approximated by the Harris expression
\begin{eqnarray} 
E_{ref} = \frac{\omega^2}{2} \Theta_0 + \frac{\omega^4}{4} \Theta_1
+ \frac{\hbar^2}{8\Theta_0}
\end{eqnarray} 
with  $\Theta_0 = 32\hbar^2$ MeV$^{-1}$ and $\Theta_1 = 32 {\hbar^4}$ MeV$^{-3}$ extracted
from the ground state band in $^{164}$Er.  The calculations are performed in the cranked shell model 
(CSM) \cite{BF.79} which employs fixed mean-field  parameters for the Nilsson potential and pairing. 
The parameters of its $E_{ref}$ reference are adjusted to the calculated Routhian of the 
$g$-configuration.  The quasiparticle orbitals, on which rotational bands are build, are
labelled by the letters of the alphabet A, B, C , D (for high-$j$ intruder states) and E, F, G, ...
(for normal parity states) [see Ref.\ \cite{F.18} for details of this labelling convention].  One can 
see that this relatively simple model can describe rather well experimentally observed features 
such as (i) large signature splitting in the pair of bands A/B, (ii) the lack or small amount of signature splitting in the 
pairs of bands E/F, X/Y and G/H, (iii) the presence of paired band crossing at $\hbar\omega_1 \approx 0.25$ 
MeV in the E/F, G/H and C bands which reflects itself in the change of the
slope of the Routhians as a function of rotational frequency  and (iv) the absence of paired band crossing in the A/B 
and X/Y pairs. The latter is due to the fact that the occupation of either of these orbitals blocks
paired band crossing in neutron subsystem. 

     Over the years huge amount of experimental data on rotating nuclei has been accumulated 
and more sophisticated theoretical tools have been developed and successfully applied for the 
description of rotating  nuclei both in paired and unpaired regimes. They go beyond the basic
assumption of the CSM on the constancy of the mean and pairing fields and take into account their 
configuration dependence either on the level of mic+mac model or on a fully self-consistent level. 
These include cranked Nilsson-Strutinsky approach  \cite{PhysRep-SBT,CRBLLP.08}, cranked 
approaches based on the Skyrme and Gogny DFTs \cite{MDD.00,ER.94,DBH.01} and CDFT 
\cite{AO.13,KR.89,ZHA.20} and others.

%%%%%%%%%%%%%%%%%%%%%%%%%%%%%%%%%%%%%%%%%%%%%%%%%%
\subsection{Single-particle and polarization effects due to the occupation of  single-particle orbitals}
%%%%%%%%%%%%%%%%%%%%%%%%%%%%%%%%%%%%%%%%%%%%%%%%%%

   The addition or removal of particle(s) to the nucleonic configuration modifies the total 
physical observables. But it also creates the polarization effects on the physical properties
(both in time-even and time-odd channels) of initial configuration. The comparison of
relative properties of two configurations can shed important light both on the
impact of the added/removed particle(s) in specific orbital(s) on physical observable of
interest and  on the related polarization effects. In addition, the comparison with experimental 
data on such properties can provide a measure of the accuracy of the description of 
single-particle properties in model calculations.

%%%%%%%%%%%%%%%%%%%%%%%%%%%%%%%%%%%%%%%%%%%%%%%%%
\begin{table}[ptb]
\caption{Experimental and calculated relative charge quadrupole moments 
$\Delta Q=Q(\mathrm{Band})-Q(^{152}\mathrm{Dy(1))}$ of the $^{149} $Gd(1), 
$^{151}$Tb(1) and $^{151}$Dy(1) superdeformed bands. The detailed structure 
of the configurations of these bands relative to the doubly magic $^{152}$Dy 
superdeformed core is given in column 2. Column 5 shows the sum 
$\sum_{i}\Delta Q^{i}$ of the 'independent' contributions $\Delta Q^{i}$ of 
$i$-th particle to the charge quadrupole moment calculated at rotational frequency $\omega=0.50$ 
MeV.  Note that the values in columns 3 and 4 are averaged over the observed
spin range. Based on the results of the CRMF calculations with the NL1 functionals 
and non-relativistic Skyrme DFT calculations with SkP and SkM* functionals
presented in Refs.\ \cite{ALR.98,SDDN.96}.}
\label{Table-qrel}
\vspace{0.5cm}
\centering
\begin{tabular}{|c|c|c|c|c|}
\hline
Band & Configuration & $\Delta Q^{exp}$ ($e$b) & $\Delta Q^{th}$ ($e$%
b) & $\sum_{i} \Delta Q^{i}$ \\ \hline
 1 & 2 & 3 & 4 & 5 \\ \hline
$^{149}$Gd(1) & $\nu[770]1/2(r=-i)^{-1} (\pi[651]3/2)^{-2}$ & $-2.5(0.3)$ & $ -2.41$ [NL1] & $-2.44$ [NL1]  \\
                        &                                                                   &                   &                        & $-2.42$ [SkP] \\
                        &                                                                   &                   &                        & $-2.32$ [SkM*] \\
$^{151}$Tb(1) & $\pi[651]3/2(r=+i)^{-1}$                              & $-0.7(0.7)$ & $-1.01$ [NL1] & $-0.99$ [NL1]  \\
                        &                                                                    &                  &                        & $-0.96$ [SkP] \\
                        &                                                                    &                  &                        & $-0.96$ [SkM*] \\
$^{151}$Dy(1) & $\nu[770]1/2(r=-i)^{-1}$                               & $-0.6(0.4)$ & $-0.53$ [NL1] & $-0.55$ [NL1]  \\
                         &                                                                    &                  &                        & $-0.57$ [SkP] \\ 
                         &                                                                    &                  &                        & $-0.48$ [SkM*] \\  \hline
\end{tabular}%
\end{table}
%%%%%%%%%%%%%%%%%%%%%%%%%%%%%%%%%%%%%%%%%%%%%%%%%%%%%%%

   Let us consider an example of rotational bands in the regime of weak pairing. 
The relative properties of different physical observables such as relative charge 
quadrupole moments
\begin{eqnarray} 
\Delta Q(\omega) = Q_B(\omega)-Q_A(\omega),
\end{eqnarray}
and relative effective alignments \cite{Rag.93} 
\begin{equation}
i_{eff}^{\,B,A}(\omega )=I_{B}(\omega )-I_{A}(\omega ),
\label{ieff}
\end{equation}%
are defined as the differences between either charge quadrupole moments $Q$ or  
the spins $I$ of the bands A and B at the same rotational frequency $\omega$
with the band $A$ being a reference band.

    They provide sensitive probes of the single-particle motion and polarization 
effects induced by the occupation of specific single-particle orbitals. In addition,  the 
experimental values of  relative  charge quadrupole moments are to a large degree free 
from the uncertainties due to stopping powers which impact the absolute values of these 
moments.  The  $\Delta Q(\omega)$ probes the differences in time-even mean fields 
generated by the addition or removal of particle(s) from the configuration of the reference
band \cite{ALR.98,SDDN.96}. 

The effective alignment $i_{eff}$ depends both on the alignment 
properties of single-particle orbital(s) by which the two bands differ and on the polarization
effects (both in time-even and time-odd channels of the DFTs) induced by the 
particles in these orbitals \cite{ALR.98}. It can also be used to investigate experimentally 
the structure of the underlying single-particle orbitals in the configurations of interest
\cite{ALR.98,Rag.93}. Note that additivity principle for the single-particle observables
has been tested for the first time on the example of effective alignments of superdeformed
bands in the $A\sim 150$ mass region \cite{Rag.93}.

%%%%%%%%%%%%%%%%%%%%%%%%%%%%%%%%%%%%%%
\begin{figure}[t!]
\centering
\includegraphics[width=8.0cm]{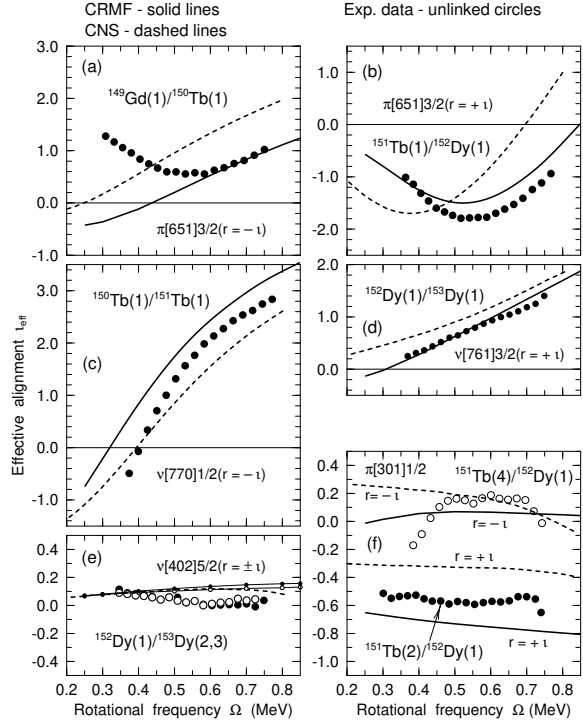}
\caption{Effective alignments, $i_{eff}$ (in units $\hbar$), extracted from experimental 
data are compared with those obtained in the CRMF and CNS calculations 
of Refs.\ \cite{ALR.98,Rag.93}.
The experimental effective alignment between bands 
A and B is indicated as `A/B'. The band A in the lighter nucleus is 
taken as a reference, so the effective alignment measures the effect 
of the additional particle. The calculated configurations differ in the 
occupation of the orbitals indicated by the quantum numbers in the 
panels. 
Figure taken from Ref.\ \cite{AR.00}.
\label{rel-alignments}
}
\end{figure}
%%%%%%%%%%%%%%%%%%%%%%%%%%%%%%%%%%%%%%%

    The comparison of experimental and calculated $\Delta Q(\omega)$ values for 
a  selected set of superdeformed bands in the nuclei near $^{152}$Dy is presented in 
Table \ref{Table-qrel} (for a more extensive set of data see Table 2 in Ref.\ \cite{SDDN.96}). 
One can see that these quantities (column 3) are well described in model calculations
(columns 4 or 5). In addition, the additivity rule of quadrupole moments \cite{SDDN.96} 
which states that  relative quadrupole moments of two configurations $\Delta Q(\omega)$ 
can be well  approximated by the sum of individual contributions  $\Delta Q^{i}$ of  
$i$-th particles to the charge quadrupole moment defined with respect of core SD configuration 
in $^{152}$Dy
\begin{eqnarray} 
\Delta Q(\omega) \approx \sum_i \Delta Q^{i}
\end{eqnarray}
is rather well fulfilled both in relativistic and non-relativistic DFT calculations (compare columns 4 
and 5 in Table \ref{Table-qrel} and see Ref.\ \cite{SDDN.96}).  This means  that the polarization 
effects due to individual particles or holes are largely independent.

    Experimental and calculated $i_{eff}$ values for single-particle orbitals active in the 
vicinity of the SD shell closures in the $A\approx 150$ region of superdeformation are 
compared in Fig.\ \ref{rel-alignments}. The calculations are carried out within the 
cranked relativistic mean field \cite{ALR.98} and the cranked Nilsson-Strutinsky
\cite{Rag.93} approaches. The CRMF calculations reproduce in average the experimental $i_{eff}$ 
values better than the CNS approach. This indicates that alignment properties of single-particle 
orbitals and their polarization effects are correctly accounted for in the CRMF approach. The discrepancy 
between  the CRMF calculations and experiment seen in Fig.\ \ref{rel-alignments}a at $\Omega\leq0.5$ 
MeV is  most likely due to the fact that pairing correlations play some role at low rotational
frequency in the $^{149}$Gd(1) band.

    It is necessary  to point on principal differences in the description of these relative 
observables in the DFT and mic+mac based approaches. The relative quadrupole moments are 
self-consistently described in the DFT based approaches \cite{ALR.98,SDDN.96,MADLN.07}. 
In contrast,  the effective single-particle quadrupole moments in the mic+mac method are not 
uniquely defined because of the lack of self-consistency between the microscopic and 
macroscopic contributions \cite{KRA.98}. In the mic+mac models  (for example, CNS) polarization 
effects caused by time-odd fields are neglected and therefore  $i_{eff}$ is predominantly defined 
by the alignment  properties of the active single-particle orbitals \cite{Rag.93}. On the contrary, 
they play an important  role in the DFT models \cite{DD.95,AR.00-to}.

   Similar studies of relative properties of rotational bands have been performed 
in different regions of nuclear chart  \cite{AF.05,MADLN.07,KRA.98}). For example, a statistical 
analysis of significant number of rotational configurations in the $A\sim 130$ region of 
superdeformation confirms the additivity principle for quadrupole moments and effective 
alignments \cite{MADLN.07}. This justifies the use of an extreme single-particle 
model in an unpaired regime typical of high angular momentum.  Note that the
basic idea behind the additivity principle for one-body operators is rooted in the independent
particle model.

%%%%%%%%%%%%%%%%%%%
\section{Conclusions}
%%%%%%%%%%%%%%%%%%%
  
  The concept of independent particle motion is the foundation of the absolute 
majority of the nuclear structure models. It leads to verifiable consequences such as shell
structure and individual single-particle properties of nucleons. Both mic+mac and DFT 
models are used nowadays for the description of numerous aspects of low energy nuclear 
phenomena.  The former allows significant flexibility (such as  the calculation of large number 
of the nucleonic configurations in a single nucleus as it is done in the CNS approach 
\cite{PhysRep-SBT}) but lacks self-consistency.  This neglect of  self-consistency limits 
the applicability of the mic+mac models to ellipsoidal  nuclei with similar density distributions 
in proton and neutron subsystems. In contrast,  the DFT models are fully self-consistent which 
makes them applicable to very exotic nuclear shapes including clustered, "bubble" and toroidal ones. 
However, the calculation of many nucleonic configurations in a given nucleus involving one or 
several blocked single-particle states still remains a challenge especially in the 
formalism which includes pairing. At present, the use of these two theoretical frameworks 
should be considered as complimentary.

   The validity of independent particle model has been confirmed by experimental
observation  of superheavy nuclei and the phenomenon of superdeformation at high spin;
these observations were motivated by model predictions. They together with global structure 
of the nuclear landscape are the consequences of underlying shell structure, which emerges from
independent particle motion of the nucleons in the nucleus. The energies and alignments of 
the single-particle states serve as a clear fingerprint of individual properties of the single-particle 
orbitals. Signature splitting properties are of particular value since either their large values 
or no splittings help in the identification of the single-particle states involved in the structure of 
observed experimental bands. The experimental features seen at extremely high spins where
the pairing correlations are negligible provide the best [at the level of single-particle states] 
examples of independent particle motion;  these are found in smooth terminating bands of the 
$A\sim 110$ mass region  and in superdeformed bands of the $A\sim 150$ mass region.

%%%%%%%%%%%%%%%%%%%
\section{Acknowledgments}
%%%%%%%%%%%%%%%%%%%

This material is based upon work supported by the U.S. Department of Energy,  
Office of Science, Office of Nuclear Physics under Award No. DE-SC0013037.
I would like to thank I. Debes and J. Dudek for the creation of improved quality
Fig.\ \ref{s-p-rot}. {Useful discussions with I. Ragnarsson are greatly appreciated.}

%%%%%%%%%%%%%%%%%%%%%%%%%%%%%%%%%
%%%   referenc.tex
%%%%%%%%%%%%%%%%%%%%%%%%%%%%%%%%%

\end{document}